\documentclass[final,5p,times,twocolumn,authoryear]{elsarticle}

\usepackage[colorlinks=true,linkcolor=blue,citecolor=blue,urlcolor=blue]{hyperref}

\usepackage{amssymb}
\usepackage{amsmath}
\usepackage{lipsum}

\usepackage{subcaption}
\usepackage[dvipsnames]{xcolor}

\newcommand{\degree}{\ensuremath{^\circ}}
\newcommand{\SC}[1]{#1}

\DeclareUnicodeCharacter{2009}{\,}
\journal{High Energy Astrophysics}

\begin{document}

\begin{frontmatter}

\title{\textbf{Effects of Coronal Mass Ejection on PSR J1022+1001 and Possible Mode Change of PSR J2145$-$0750 in the InPTA DR2}}

\author[IMSc]{Shaswata Chowdhury\corref{cor1}}
\ead{shaswata.phyres@gmail.com}
\cortext[cor1]{Corresponding author}
\author[RAC]{M. A. Krishnakumar}
\author[IMSc,HBNI]{Manjari Bagchi}
\author[NCRA,IITR]{Bhal Chandra Joshi}
\author[Kumamoto]{Nobleson K.}
\author[IITI]{Jibin Jose}
\author[IITH]{Shantanu Desai}
\author[IISERM]{Manpreet Singh}
\author[IITH]{Vaishnavi Vyasraj}
\author[CEBS]{Kuldeep Meena}
\author[RRI]{Amarnath}
\author[IITI]{Manoneeta Chakraborty}
\author[IMSc]{Shubham Kala}
\author[NWU]{Debabrata Deb}
\author[IISc]{Zenia Zuraiq}
\author[RRI,Christ]{Arul Pandian B.}
\author[IITD]{Neelam Dhanda Batra}
\author[PRL]{Churchil Dwivedi}
\author[IMSc,HBNI]{Sushovan Mondal}
\author[IISc]{Avinash Kumar Paladi}
\author[IISERB]{Kaustubh Rai}
\author[AEI]{Abhimanyu Susobhanan}
\author[IITR]{Adya Shukla}
\author[GLA,IITH]{Aman Srivastava}
\author[IISERB]{Mayuresh Surnis}
\author[IISERB]{Hemanga Tahbildar}
\author[Kumamoto]{Keitaro Takahashi}
\author[INAF]{Pratik Tarafdar}
\author[RRI]{Prabu Thiagaraj}
\author[IISERB]{Kunjal Vara}

\address[IMSc]{The Institute of Mathematical Sciences, C. I. T. Campus, Taramani, Chennai 600113, India}
\address[RAC]{Radio Astronomy Centre, National Centre for Radio Astrophysics, Tata Institute of Fundamental Research, Udhagamandalam, 643001, India}
\address[HBNI]{Homi Bhabha National Institute, Training School Complex, Anushakti Nagar, Mumbai 400094, India}
\address[NCRA]{National Centre for Radio Astrophysics, SP Pune University Campus, Pune, Maharashtra, 411007, India}
\address[IITR]{Department of Physics, Indian Institute of Technology Roorkee, Roorkee, Uttarakhand, 247667, India}
\address[Kumamoto]{Faculty of Advanced Science and Technology, Kumamoto University, 2-39-1 Kurokami, Kumamoto 860-8555, Japan}
\address[IITI]{Department of Astronomy, Astrophysics, and Space Engineering, Indian Institute of Technology Indore, Indore 453552, India}
\address[IITH]{Department of Physics, IIT Hyderabad, Kandi, Telangana 502284, India}
\address[IISERM]{Department of Physical Sciences, Indian Institute of Science Education and Research (IISER) Mohali, Sector 81, SAS Nagar, Mohali, Punjab, 140306, India}
\address[CEBS]{UM-DAE Centre for Excellence in Basic Sciences, University of Mumbai, Vidyanagari, Mumbai 400098, India}
\address[RRI]{Raman Research Institute, Bengaluru 560080, Karnataka, India}
\address[NWU]{Centre for Space Research, North-West University, Private Bag X6001, Potchefstroom 2520, South Africa}
\address[IISc]{Department of Physics, Indian Institute of Science, C. V. Raman Avenue, Bengaluru 560012, India}
\address[Christ]{Department of Physics and Electronics, The Christ (Deemed to be University), Bangalore, India}
\address[IITD]{Department of Physics, Indian Institute of Technology Delhi, Hauz Khas, New Delhi 110016}
\address[PRL]{Astronomy and Astrophysics Division, Physical Research Laboratory, Thaltej Campus, Ahmedabad 380059, Gujarat, India}
\address[IISERB]{Department of Physics, IISER Bhopal, Bhauri Bypass Road, Bhopal, 462066, India}
\address[AEI]{Max Planck Institute for Gravitational Physics (Albert Einstein Institute), Callinstraße 38, D-30167 Hannover, Germany}
\address[GLA]{Department of Physics, GLA University, Mathura 281406, India}
\address[INAF]{INAF - Osservatorio Astronomico di Cagliari, via della Scienza 5, 09047 Selargius (CA), Italy}

\begin{abstract}
The Indian Pulsar Timing Array (InPTA) has recently published its second data release (DR2), comprising the timing analysis of seven years of data on 27 millisecond pulsars (MSPs), observed simultaneously in the 300$-$500 MHz (band 3) and 1260$-$1460 MHz (band 5), using the upgraded Giant Metrewave Radio Telescope (uGMRT). The low-frequency data, particularly in band 3, is highly sensitive to propagation effects such as dispersion measure (DM) fluctuations, which can be imprints of some astrophysical phenomena (scientific outliers). Here, we analyze the two outliers of possible astrophysical origin coming from the band 3 DM time series of two pulsars: PSR J1022+1001, with an ecliptic latitude of $-0.06^{\circ}$, and PSR J2145$-$0750, one of the brightest MSPs, with multi-component profile morphology. Our study reveals compelling evidence for a coronal mass ejection (CME) event traced in the data of PSR J1022+1001, and reports evidence for a potential mode-changing event in PSR J2145$-$0750. \SC{By contrasting these two cases, we show that DM fluctuations due to CME interacions and intrinsic mode-changing events produce distinct observational signatures, enabling a physically informed classification of scientific outliers in PTA datasets.} Extending the analyses presented here to the full sample of InPTA-DR2 pulsars is expected to reveal additional CME events, and possible mode-changing events. Such detections will not only improve our understanding of solar and pulsar magnetospheric plasma interactions but will also enable more accurate modelling of DM variations, leading to improved pulsar timing solutions, which are crucial for high-precision Pulsar Timing Array (PTA) science.
\end{abstract}

\end{frontmatter}

\section{Introduction}

High-precision timing of radio pulsars in applications such as the search for low-frequency gravitational waves (GWs) using pulsar timing array experiments (PTAs) often faces the challenge of dealing with shifts in time-of-arrival (ToA) of the radio pulse~\citep{Verbiest24}. Such shifts can be caused by variations 
in the propagation delay through the ionized inter-stellar medium (IISM) -- extreme scattering events~\citep{extreme_scat}, scintillations~\citep{scintillation_lowfreq, scintillation_EPTA}, as well as through interplanetary medium (IPM) -- coronal mass ejection (CME)~\citep{Howard_CME}, solar wind (SW)~\citep{Caterina_SW_PTA_1, Caterina_SW_PTA_2, Caterina_SW_PTA_3}, being a few of the explored phenomenons.
Such shifts can also be caused by an intrinsic change in the pulse profile from the usually stable \SC{template profile (see~\citet{Rana25} for the latest template generation scheme)}, due to glitches~\citep{himanshu_glitch_1, himanshu_glitch_2}, magnetospheric origins~\citep{Rami_profilechange_J1713}, or mode change~\citep{modechange_J1909_Miles}. This reveals a systematic shift in the 
timing residual obtained after a comparison of the observed ToA with 
that predicted by a model of the pulsar incorporating astrometric, spin, 
propagation, and binary parameters. Therefore, these ``outliers'' 
may indicate astrophysical events of interest, which may be one-off 
transient events or longer time scale events. A couple of good examples are the profile change event in PSR J1713+0747 reported in 2018~\citep{Lam_J1713_Profilechange}, in 2020~\citep{Goncharov_J1713_ProfileChange}, and in 2021~\citep{Singha21} (subsequent interpretations can be found in~\citet{Jennings_J1713_profilechange, Rami_profilechange_J1713}), and the CME-event reported in PSR J2145$-$0750~\citep{KK_CME_DMCalc}. 

These \SC{astrophysical phenomena} can affect the estimated dispersion measure \SC{(DM)}\footnote{\SC{It is a measure of the propagation delay, proportional to the integrated column density of electrons in the inter-stellar and inter-planetary medium along the line-of-sight to the pulsar.}} of the pulsars. Finding and understanding the causes of such outliers in the  DM time series is very important. First, it might lead to serendipitous discovery of interesting astronomical events like profile shape change of the pulsar. Second, it has the potential to improve our understanding of how different astrophysical phenomena impact each other, e.g., CME changing the DM of a pulsar. 
Such outlier events could also affect noise modelling~\citep{Aman23} and degrade the \SC{sensitivity to gravitational wave searches}~\citep{Golam_Outliers_GWB}.
Such studies from one pulsar timing experiment (the Indian pulsar timing array (InPTA)~\citep{Joshi18, Joshi22} in the present case) can potentially alert other PTAs to investigate their data, following the example of the profile change event on PSR J1713+0747. As PTA data can be used not only for the detection of low-frequency GWs, but also be used for various other purposes, like improving solar system ephemeris~\citep{SolarSystem_PTA_Champion2010, SolarSystem_PTA_Caballero2018, SolarSystemEphemerisError_PTA_Li2016, SolarSystemLINIMOSS_PTA_Guo2019},  testing theories of gravity~\citep{gravitytest_PTA_1, gravitytest_PTA_2, gravitytest_PTA_3, gravitytest_PTA_4}, improving the mass measurements of the neutron stars~\citep{neutronstarmass_PTA_1} and much more, understanding of outliers in the PTA data is always necessary.

Such outliers can easily be determined through the decade-long monitoring 
of millisecond pulsars (MSPs) in the PTA experiments. Currently, six  such experiments are ongoing with time baselines ranging from 4 years 
to 28 years (European Pulsar Timing Array (EPTA)~\citep{EPTA}, the InPTA~\citep{Joshi18,Joshi22}, the Parkes Pulsar Timing
Array (PPTA)~\citep{PPTA}, North American Nanohertz Observatory for Gravitational Waves
(NANOGrav)~\citep{McLaughlin2013},
Meerkat Pulsar Timing Array~\citep{MPTA} and the Chinese Pulsar Timing Array (CPTA)~\citep{CPTA2025}). The 
data and resources from these PTA experiments are pooled together by 
the International Pulsar Timing Array (IPTA)~\citep{IPTA} to form the most sensitive 
data set for the search for GWs. These experiments have 
reported evidence for the low frequency GWs~\citep{Agazie2023,epta+inpta2023a,CPTAGW,MPTA,PPTAGW}.

Recently, InPTA made its second data release (DR2) public~\citep{Rana25} 
with seven years of data on 27 MSPs. This work investigates outliers in the DM time-series of two pulsars from the InPTA-DR2 sample and evaluates their potential astrophysical origin. The two pulsars of interest, highlighted here, are -- (a) PSR J1022+1001, which  has a very low ecliptic latitude of $-$0.06$\degree$, and is known for exhibiting significant solar wind-induced DM variability due to its close proximity to the Sun, making it a natural candidate for probing solar events such as CMEs; (b) PSR J2145$-$0750, one of the brightest pulsars in the InPTA sample, shows pronounced short-timescale variations in its DM time series, and its multi-component profile morphology allows sensitive tests for intrinsic mode-changing behavior. \SC{Distinguishing the deviation in DM arising from CME interactions, and from intrinsic 
mode changes, requires that the very definition of an outlier be tailored to the physical scenario under consideration. In this work, we adopt such a scenario-driven approach: the definition of an outlier for PSR J1022+1001 is based on solar elongation geometry and deviations from modelled solar-wind DM, whereas for PSR J2145$-$0750 it is based on statistically significant changes in its pulse-shape properties. This framework is essential for isolating astrophysically motivated events in PTA datasets, before applying deeper diagnostic analyses to confirm their origin.}

The outline of this paper is as follows. We provide an overview of the observational strategy of InPTA-DR2 and the analysis carried out in this paper, in Section \ref{sec:Obs}. In Section \ref{sec:CME} the investigation for the detection of CME using PSR J1022+1001 data is demonstrated, while in Section \ref{sec:ModeChange} we describe the analysis for a potential mode change in PSR J2145$-$0750. \SC{Section \ref{sec:compare} provides a comparative discussion of these two case studies, highlighting the observational signatures that distinguish CME-driven and mode-change-driven DM excursions.} The implications of the results and future directions are discussed in Section \ref{sec:conclude}. 

\section{Observations and Analysis}
\label{sec:Obs}

The InPTA has been monitoring 27 MSPs routinely using simultaneous observations 
in the 300$-$500 MHz (band 3) and 1260$-$1460 MHz (band 5) \SC{bands,} \SC{as well as single band 3 observations at low frequency, all} with a bandwidth of 200 MHz\footnote{\SC{Band 1 (50-80 MHz) is not yet commissioned by uGMRT, while band 2 (120-250 MHz) is unsuitable for high-precision PTA science due to strong scattering and substantial radio frequency interferences (RFIs). Initial tests showed that band 4 (550–750 MHz or 650-850 MHz) observations in triple band or dual band configuration could not offer any better sensitivity or DM precision, and hence not considered.}}. The unique subarray capability of the upgraded Giant Metrewave Radio Telescope (uGMRT)~\citep{Gupta17} 
is utilized by InPTA to achieve this concurrency. The uGMRT antennae 
are split into two phased arrays with 10 and 15 antennae in band 3 and 5 respectively\SC{, for band 3-5 dual band observation, while a phased array of 20 antennae are used for the single band 3 observations.} \SC{The data were recorded with the coherent dedispersion~\citep{CDP_Yashwanth} mode at band 3 and using incoherent
dedispersion at band 5\footnote{\SC{Since the low frequency, band 3, is strongly affected by dispersive effects, even a small width of the frequency channels retains small amount of dispersion effect, that remains uncorrected in the incoherent dedispersion used. Hence the real time coherent dedispersion is used to mitigate the limitation due to finite width of the frequency channels in band 3. However, since the high frequency, band 5, is not significantly affected by dispersion, incoherent dedispersion is sufficient to correct for dispersive delays in band 5.}}.}
High precision epoch-wise 
DM time series is 
obtained by the InPTA with these data. \SC{It is notable that} InPTA is the only 
PTA experiment which is capable of obtaining instantaneous DM measurements using 300 
to 1500 MHz concurrent wide-band observations, unlike other PTA experiments, which utilize 
multiple epochs to carry out low and high frequency observations.
\SC{However, the single band 3 observation data has a relatively higher baseline compared to the dual band 3-5 configuration, and since the lowest frequency band 3 is most sensitive to propagation effects, we work with only band 3 observation data here. Besides, since any intrinsic changes to the pulsar is expected to leave a broad band (in frequency) signature, we should, in principle, be able to capture such intrinsic changes as well in our single band 3 data.}  
In this paper, we use the intermediate and final products of InPTA-DR2 for further investigations. The observing setup, observing 
strategy and data processing used in InPTA are described in~\citet{Rana25, Abhimanyu_PINTA, Tarafdar22}, where the reader can find additional details.

\SC{The starting point of our analysis, in this paper, is the intermediate epoch-wise DM estimates, along with their associated uncertainties\footnote{\SC{DM uncertainty refers to the confidence interval on our {\tt DMCalc}-evaluated DM value. In particular, the DM uncertainty is the 1-$\sigma$ formal error returned by a weighted least-squares fit in {\tt TEMPO2}~\citep{Hobbs_TEMPO2,Edwards_TEMPO2}, performed on the cleaned subbanded TOAs -- after removal of low-S/N points and Huber-regression-based outliers as implemented in {\tt DMCalc}.}}, in band 3, produced during the InPTA-DR2 analysis, using {\tt DMCalc}~\citep{KK_CME_DMCalc}. Such DM time series contains fluctuations arising from a mixture of propagation effects, and intrinsic pulsar variability. Moreover, because different physical processes imprint qualitatively different signatures on the DM series, defining an outlier purely from a generic statistical deviation is insufficient and may even be misleading. Instead, the astrophysically motivated definition of an outlier is tailored to the physical scenario under investigation, as will be explained in the following sections. In addition to these astrophysical reasons, apparent DM deviations may also arise from observational systematics; however, poor signal-to-noise ratio (S/N) epochs affected by instrumental issues, or other data-quality concerns are systematically excluded through standard post-processing diagnostics and observatory logs, ensuring that only genuine astrophysical candidates are considered. This contextual criterion for defining outliers is crucial: it allows us to isolate physically meaningful events from the DM series before performing the deeper diagnostic analyses that ultimately confirm their astrophysical origin.} 

The candidate outlier for PSR J1022+1001 is cross-checked against solar probe data, coronagraph images, and solar wind simulations. Consistency between enhanced electron densities, CME trajectories, and the pulsar’s line of sight is used to validate the association. For PSR J2145$-$0750, \SC{the candidate outlier is} studied in frequency-resolved folded profiles (300$-$500 MHz), allowing us to establish if the observed profile departs from the standard template in a manner consistent with mode change. This targeted approach enables us to highlight one astrophysically significant outlier for each of the two pulsars, avoiding the ambiguity of treating all statistical outliers equally. We also briefly examined whether any of \SC{these} DM outliers could be influenced by terrestrial effects such as lightning or terrestrial gamma-ray flashes (TGFs), which are known to cause short-lived ionospheric perturbations \citep{kuo2015ionospheric} and radio emissions around 300--400 MHz \citep{richard1986results}, overlapping with our observing band. No consistent correlations were found, suggesting that  these effects do not significantly impact our observing bands of interest. In the following sections, we explain the methodology and results in detail for the two pulsars of interest.

\section{Investigating the effect of CMEs on PSR J1022+1001}
\label{sec:CME}
A CME is a large-scale eruption of magnetized plasma from the Sun’s corona that expels bulk of charged particles into interplanetary space. To assess whether \SC{any observational epoch of PSR J1022+1001 is affected by such a solar energetic event, we examine in a step-wise manner.} We first determine\SC{, from the DM time series,} the outlying epochs most likely to be affected by CMEs, then identify candidate CME events from solar monitoring databases. Next, we check for corroborating signatures in in situ solar wind measurements, before finally estimating the expected DM excess contributed by such an event. This structured approach \SC{as detailed in the following subsections,} allows us to systematically test the solar origin of the observed anomaly.

        \subsection{Identifying the possible CME affected observations}
        \label{sec:cme_affected_epoch}
        \SC{For a low-ecliptic-latitude pulsar such as PSR J1022+1001, where solar-wind variability and CME encounters are expected, a solar-event-induced outlier is naturally defined as a DM enhancement that exceeds the predictions of a solar-wind model within a small solar elongation. This restriction to small elongations is motivated by the fact that, the pulsar signal traverses denser and more variable heliospheric plasma, making any CME-related DM excursion physically identifiable.} 
        
\begin{figure}[htbp]
                    \centering
                    \includegraphics[width=0.45\textwidth]{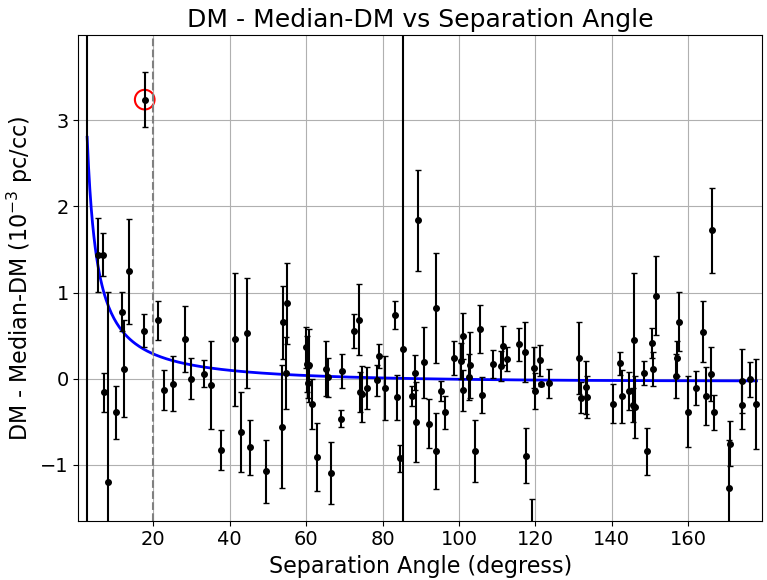}
                    \caption{`Observational' DM and `modelled' DM vs separation angle (with respect to Sun) for PSR J1022+1001, with the outlier of interest being marked with red circle. The black points \SC{with vertical bars represent `observational' DM with its uncertainty}, while the blue curve denotes the `modelled' DM from the spherically symmetric solar wind model.}
                    \label{fig:J1022_dmexcess}
                \end{figure}

        \SC{Therefore,} we first determined the angular separation between the Sun and the pulsar’s line of sight at all observing epochs. \SC{We} subtracted the median DM from the epoch-wise DM values -- we will refer to this difference as `observational' DM from here onwards. We then examined the variations in observational DMs as a function of solar elongation. In parallel, we estimated the DM contribution expected from the widely used spherically symmetric solar-wind model~\citep{Caterina_SW_PTA_3} using the $NE\_SW$, parameter obtained from the timing solution \SC{of InPTA-DR2}~\citep{Rana25}. This DM is subtracted from its median across the DR2 baseline -- we refer to this difference as `modelled' DM. We refer to the difference between the `observational' and 'modelled' DM as the DM excess. \SC{In Fig.~\ref{fig:J1022_dmexcess} the black points with vertical bars represent the observational DM with its uncertainty, while the modelled DM is represented by the blue curve.}
                
        We flagged \SC{as outliers those observational epochs within a conservative $20\degree$ solar elongation threshold for which the observational DM exceeds the modelled DM by more than 3 times the corresponding DM uncertainty.} \SC{For PSR J1022+1001, this procedure yields a promising candidate outlier -- MJD 59800 (09 August 2022), with a separation angle of 17.75\degree and a DM excess of $2.91\times10^{-3}$ pc/cc (see Fig.~\ref{fig:J1022_dmexcess}).} The profile for this epoch has \SC{reasonably high S/N with good detection -- }no visible deviation from its template profile -- thereby ruling out any \SC{instrumental or} intrinsic origin for such DM excess.

        \subsection{CME responsible for the DM outlier}
        To identify the most plausible CME event(s) that could give rise to such DM excess, we consulted the DONKI\footnote{The Space Weather Database Of Notifications, Knowledge, Information (DONKI), developed at the Community Coordinated Modeling Center (CCMC), is a comprehensive on-line tool for space weather forecasters, scientists, and the general space science community -- \url{https://kauai.ccmc.gsfc.nasa.gov/DONKI/}.}~\citep{DONKI_1} database for solar eruptive events for three days prior to MJD 59800. From this list, we assess which modelled CMEs are likely to intersect the pulsar's line of sight (LOS), and whether they are simultaneously accessible to any spacecraft capable of providing in situ plasma measurements. Two CME events -- cme1, and cme2, emerge as relevant candidates. Both were detected by the Solar TErrestrial RElations Observatory (STEREO)-A spacecraft~\citep{STEREO_Kaiser_1}, which at the time was located at 0.95 AU from the Sun with a Heliocentric Earth Ecliptic (HEE) latitude of 0.12\degree and longitude of $-$21.97\degree. The main properties of these events are summarized in Table~\ref{tab:cmeeventsJ1022}.
          \begin{table}[htbp]
                \centering
                \begin{tabular}{|c| c| c|}
                \hline
                \multicolumn{1}{|c|}{CME} & \multicolumn{1}{c|}{cme1} & \multicolumn{1}{c|}{cme2} \\
                \hline
                \multicolumn{1}{|c|}{lat} & \multicolumn{1}{c|}{$-$23.0\degree} & \multicolumn{1}{c|}{$-$33.0\degree} \\
                \hline
                \multicolumn{1}{|c|}{long} & \multicolumn{1}{c|}{$-$44.0\degree} & \multicolumn{1}{c|}{$-$35.0\degree} \\
                \hline
                \multicolumn{1}{|c|}{speed (km/s)} & \multicolumn{1}{c|}{912} & \multicolumn{1}{c|}{776} \\
                \hline
                \multicolumn{1}{|c|}{half-width} & \multicolumn{1}{c|}{39.0\degree} & \multicolumn{1}{c|}{20.0\degree} \\
                \hline
                \multicolumn{1}{|c|}{start-time (UT)} & \multicolumn{1}{c|}{06-08-2022T01:48} & \multicolumn{1}{c|}{07-08-2022T11:00} \\
                \hline
                \end{tabular}
                \caption{Properties of CMEs that could have affected PSR J1022+1001 observation. The latitude (lat) and longitude (long), in Stonyhurst heliographic coordinate system, represent the direction along which the nose of the individual CMEs propagate. These CME parameters are taken from the DONKI database.}
                \label{tab:cmeeventsJ1022}
            \end{table}
            
\begin{figure}[htbp]
                \centering
                \includegraphics[width=0.48\textwidth]{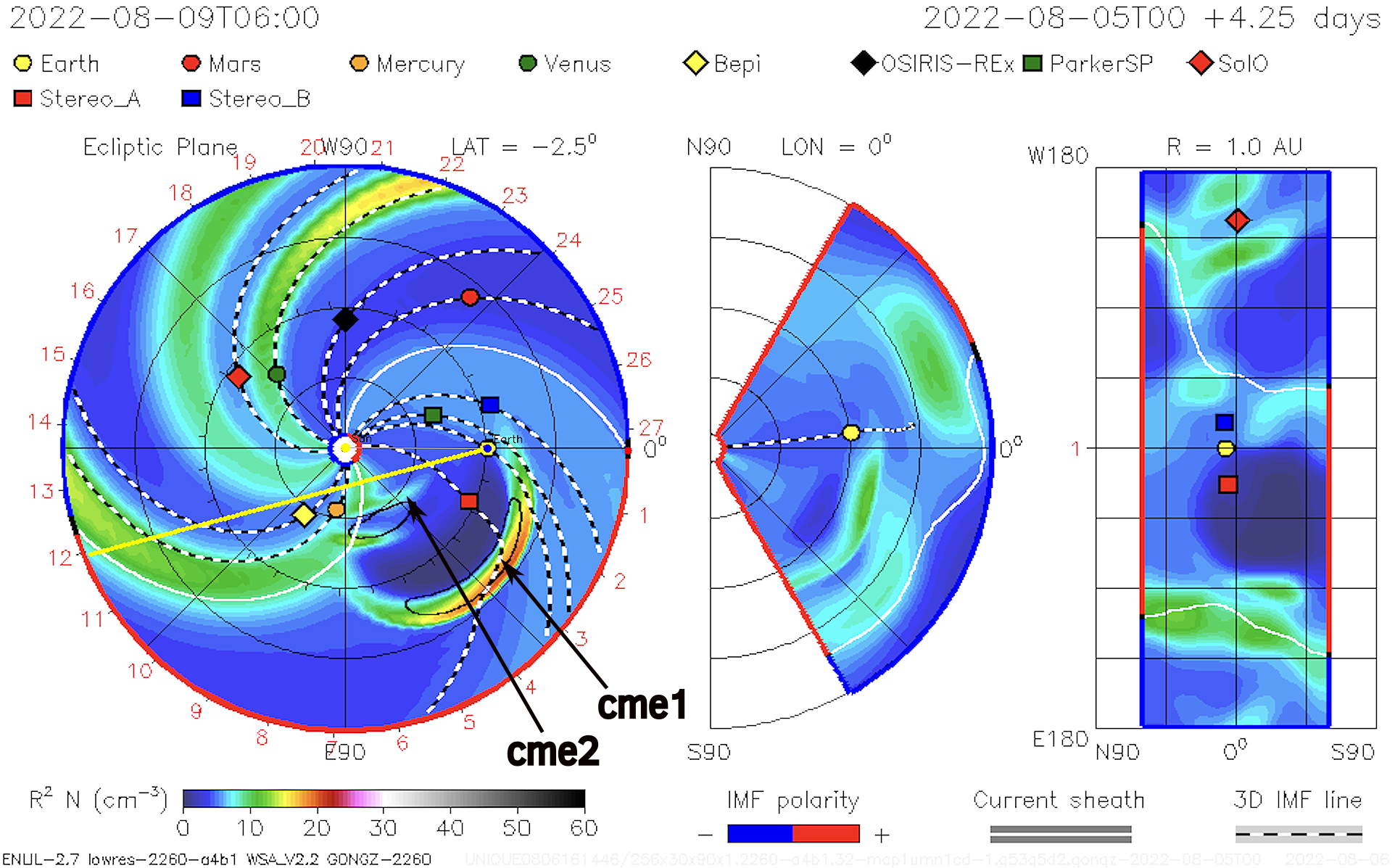}
                \caption{PSR J1022+1001's LOS (yellow line) overlaid on the MHD simulation snapshot taken at 06:00:00 (UTC), around the pulsar observation time on 09 August 2022. Locations of various planets and satellites are shown with different symbols as explained in the legends. The dashed spirals are the \SC{Parker spirals}~\citep{ParkerSpiral1958}, the colour code represent the density as per the colorbar. The two CMEs -- cme1, and cme2 are also labelled. These snapshots are taken from the MHD simulations in DONKI.}
                \label{fig:J1022_LOS_overlaid}
            \end{figure}

        To further evaluate the influence of these CMEs on the pulsar's LOS, we turn to visualizations of the DONKI simulations performed within the WSA–ENLIL+Cone (WEC) framework~\citep{WSA, ENLIL, WSA_ENLIL_CONE}. Fig.~\ref{fig:J1022_LOS_overlaid} shows the LOS of PSR J1022+1001 overlaid on the simulation snapshot at 06:00:00 (UTC) on 09 August 2022 (the CME-affected epoch). 
        Given the 6-hour temporal resolution of the simulation, which is much longer than the 17-minute observation duration of the pulsar on that day, this snapshot corresponds to the closest available time to the actual observation time (05:18:39 UTC) on 09 August 2022. The figure displays the positions of the Sun, Earth, satellites, LOS of the PSR J1022+1001, all projected onto the ecliptic plane -- the Sun is marked with a white circle at the centre, while the Earth is marked with a yellow circle, 1 AU away from the centre, along the $0\degree$ longitude; the projection of the LOS of PSR J1022+1001 is shown with the yellow line, while those for the satellites and planets are shown with varying colours. The first CME (cme1), associated with a dense, fast outflow, passes beyond both the LOS and STEREO-A, as shown in Fig.~\ref{fig:J1022_LOS_overlaid}. 
       The second CME (cme2), however, marginally intersects the LOS during the InPTA observation window, and may represent either a compressed solar wind structure trailing cme1 or a weaker independent event. In either case, it is cme2 that is temporally coincident with the anomalous DM behaviour, because of its influence on the pulsar's LOS on the observation time. The position of the satellites are obtained from the STEREO Science Center, maintained by the National Aeronautics and Space Administration (NASA)\footnote{ \url{https://stereo-ssc.nascom.nasa.gov/cgi-bin/make_where_gif}}.

\begin{figure}[htbp]
                \centering
                \includegraphics[width=0.48\textwidth]{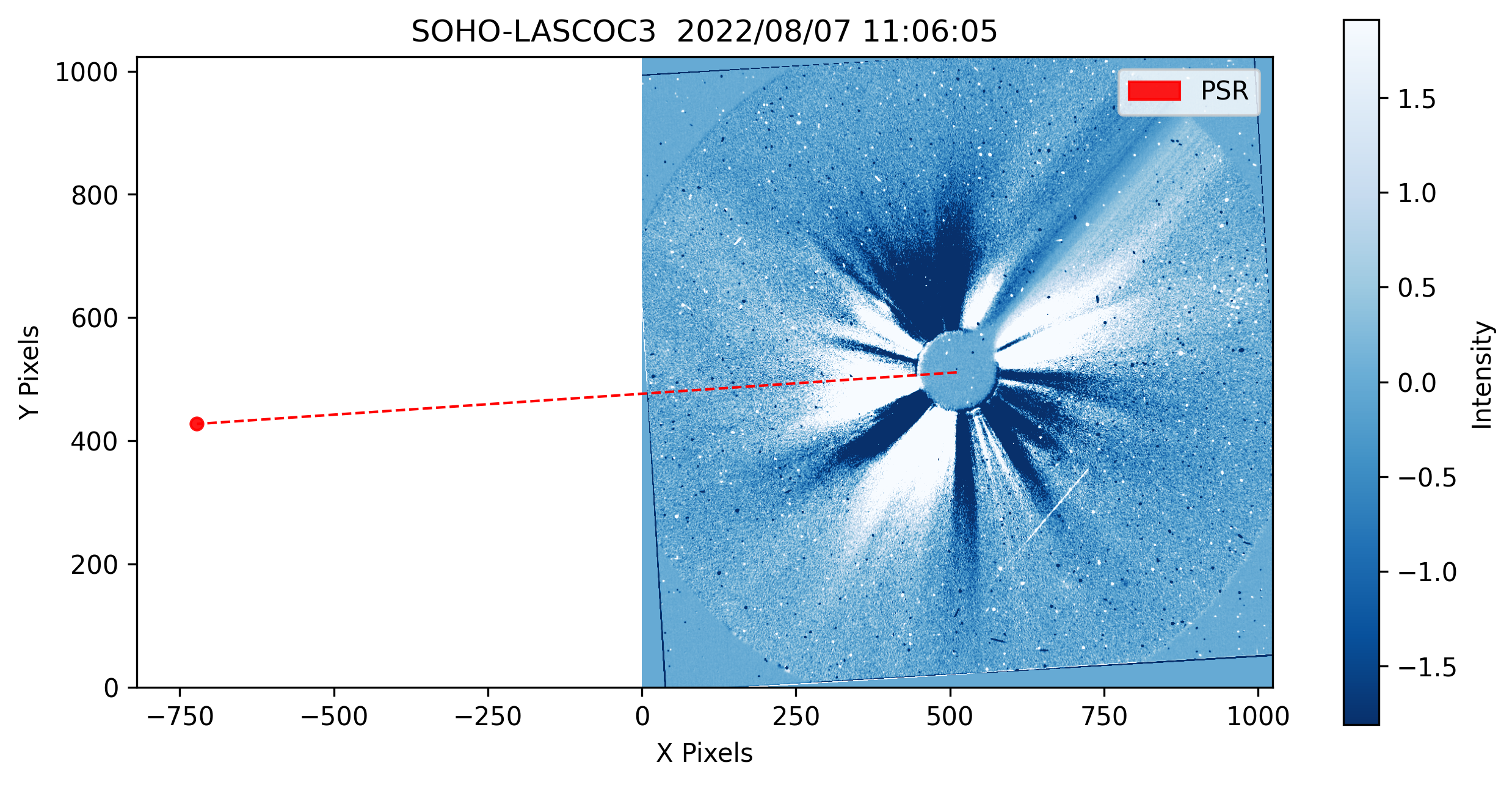}
                \caption{LASCO/C3 coronagraph base difference image plotted from the fits file corresponding to the date of cme2, which passed through the LOS of PSR J1022+1001.}
                \label{fig:J1022_LASCO_C3}
            \end{figure}

        Additional support comes from the Large Angle and Spectrometric Coronagraph – C3 (LASCO/C3 coronagraph)~\citep{LASCO_C3} data. The base-difference images generated from the corresponding FITS files are carefully oriented with solar north upwards and the Sun centered in the frame. These reveal large-scale outflows associated with the CME events. For PSR J1022+1001, the CME event of interest, i.e., cme2 manifests as a clear mass outflow aligned toward the pulsar LOS \SC{(see Fig.~\ref{fig:J1022_LASCO_C3}), where }the pulsar is marked as a red dot, with a dotted line indicating the direction in which the CME should travel to influence the pulsar's LOS. The LASCO/C3 field of view extends to $\sim$ 8$\degree$ from the Sun’s center; PSR J1022+1001 lies at an angular separation of 17$\degree$, and hence falls outside the image plane. Nevertheless, the inferred trajectory of the CME is consistent with an impact on the LOS.

        \subsection{In situ data}
\begin{figure*}[t]
                \centering
            
                \begin{subfigure}[b]{0.45\textwidth}
                    \centering
                    \includegraphics[width=\textwidth]{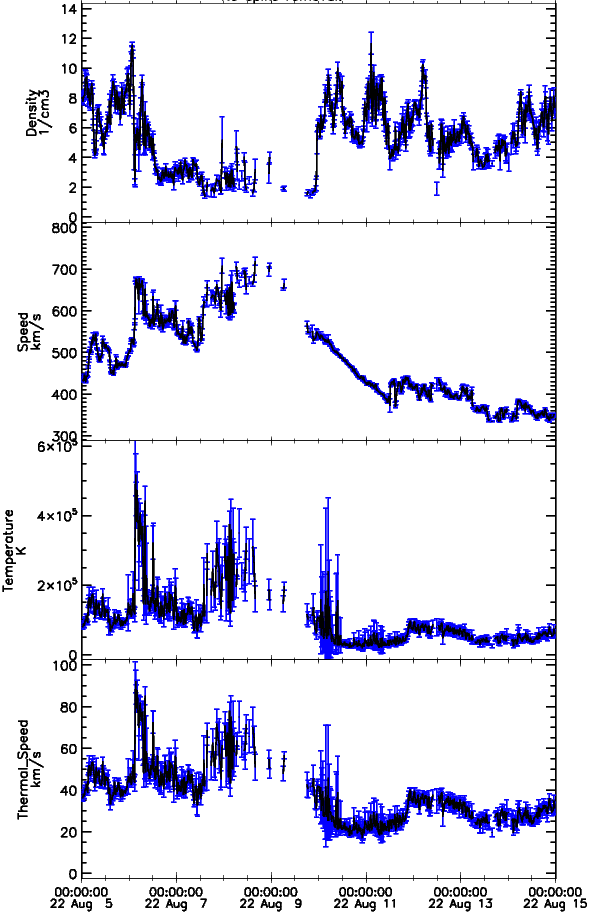}
                    \label{fig:SA_density}
                \end{subfigure}
                \hspace{0.03\textwidth}
                \begin{subfigure}[b]{0.45\textwidth}
                    \centering
                    \includegraphics[width=\textwidth]{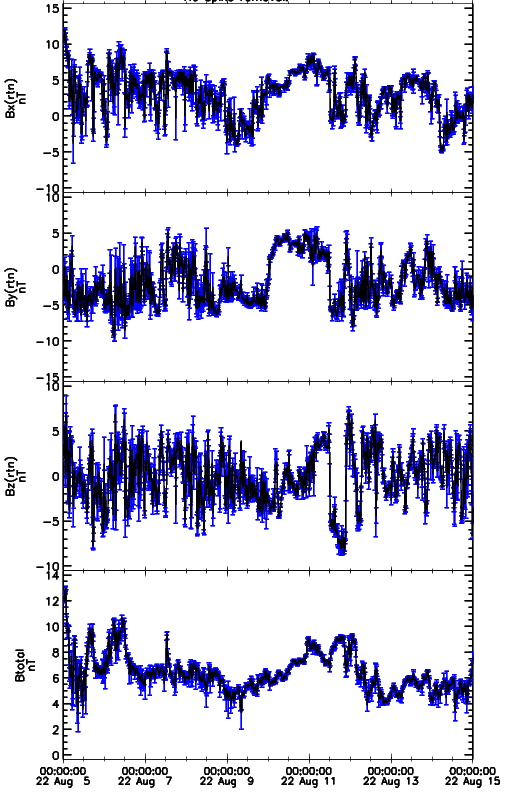}
                    \label{fig:SA_mag}
                \end{subfigure}
            
                \caption{In situ data from STEREO-A for a span of 10 days around the cme2 event is shown. The left panel shows the plasma density  in cm$^{-3}$, SW speed, Temperature, and the thermal speed. The right panel shows the measured magnetic field in X,Y and Z directions as well as  the total value,  in units of nT. These plots are generated using the CDAWeb.}
                \label{fig:SA_insitu}
            \end{figure*}

            The in situ data from STEREO-A reveals that the cme1, after starting from the Sun at 06-08-2022 01:48, with a speed of 900 km/s reaches STEREO-A around 08-08-2022 00:00 UTC; we see an enhancement around that time in the measured speed and temperature of the solar wind (see Fig.~\ref{fig:SA_insitu} - generated using CDAWeb\footnote{Coordinated Data Analysis Web (CDAWeb) is part of NASA’s Space Physics Data Facility (SPDF), which serves as a central archive and access point for heliophysics and space weather data from various NASA and international missions - \url{https://cdaweb.gsfc.nasa.gov/}}). 
           \SC{The data coverage is incomplete for unknown reasons, which may explain the absence of the expected plasma-density jump}. Nevertheless, we see subsequent increases in the plasma density after 10-08-2022 00:00 UTC, which notably corresponds to the cme2, which after starting from the Sun at 07-08-2022 11:00 UTC, with a speed of 776 km/s is supposed to reach STEREO-A around 09-08-2022 17:00 UTC. The differences \SC{between the predicted and observed arrival times of the CME likely arises from simplistic assumptions, such as constant CME velocity,} which is not completely valid given the fact that CME can accelerate or decelerate along its path due to the drag force~\citep{CME_SW_drag_Maloney_1, CME_SW_drag_2, CME_SW_drag_3}. The second CME (cme2) seems to be a possible secondary CME coming from nearly the same active region (as cme1, see Table~\ref{tab:cmeeventsJ1022}), that got accelerated by possible interaction with the fast solar wind component.

        \subsection{DM excess estimate}
\begin{figure*}[htbp]
            \centering
        
            \begin{subfigure}[b]{0.4\textwidth}
                \centering
                \includegraphics[width=\textwidth]{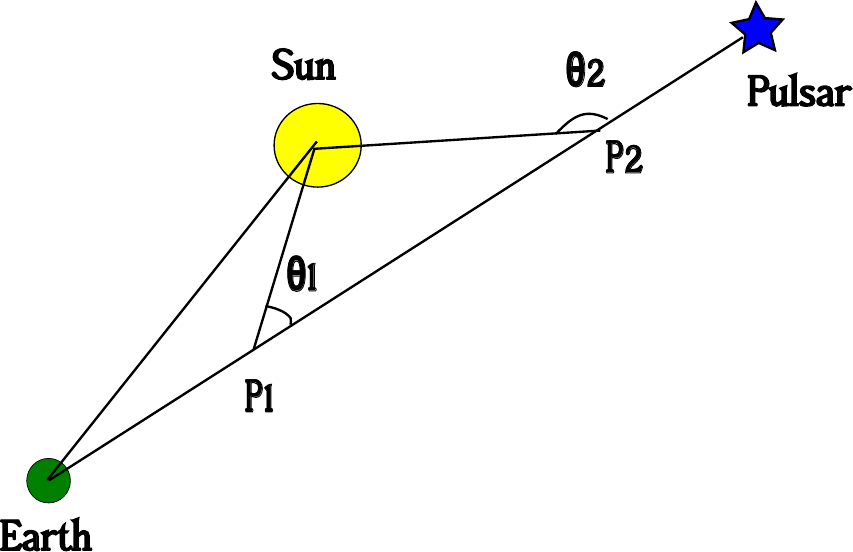}
                \caption{Schematic illustration of integrating-path for DM evaluation}
                \label{fig:dm_illustration}
            \end{subfigure}
            \hspace{0.05\textwidth}
            \begin{subfigure}[b]{0.4\textwidth}
                \centering
                \includegraphics[width=\textwidth]{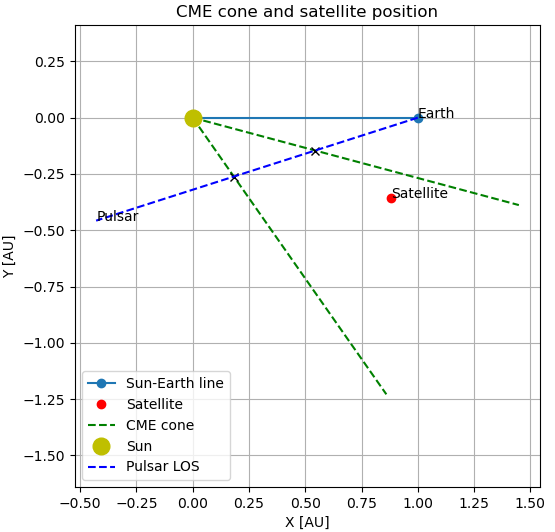}
                \caption{Propagation path of CME affecting the LOS for PSR J1022+1001}
                \label{fig:cme_impact_J1022}
            \end{subfigure}
        
            \caption{Geometry of paths: (a) DM estimation and (b) impact of CME on LOS of PSR J1022+1001}
            \label{fig:geometry}
        \end{figure*}

        Here, we estimate the DM excess from the in situ data and the spherically symmetric SW model. From a spherically symmetric SW model consideration, the formula for the DM contribution coming from a portion of the LOS, say $P_1$-$P_2$ (see Fig.~\ref{fig:dm_illustration}) is
        \begin{equation}
            DM_{P_1-P_2} = \frac{n_{sat} (r_{sat} (\text{AU}))^2}{1~ \text{AU}}\int_{\theta1}^{\theta2}d\theta,
        \label{eq:DM_excess}
        \end{equation}
        where $n_{sat}$ is the electron density at the position of the satellite of interest, $r_{sat}~(\text{AU})$ is the distance of the satellite from the Sun in units of AU, $\theta_i$ is the angle subtended by the sun-$P_i$ line on the earth-pulsar LOS, where $i\in\{1,2\}$.\\

        Using the position coordinates of the satellite of interest, the pulsar's LOS, and the properties of the CME modelling as per DONKI, we identify the portion of the LOS that gets influenced by the cme2's impact on the day of our CME-affected-observation \SC{(see Fig.~\ref{fig:cme_impact_J1022})}.
        
            The \SC{value of} $n_{sat}$ \SC{corresponding to cme2, as measured} from the in situ data of STEREO-A, is roughly 10~cm$^{-3}$.  Integrating along the affected portion of the LOS, in Eq. \ref{eq:DM_excess}, we get the DM due to both the CME and the background SW $\sim$ $5.7\times10^{-3}$ pc~cm$^{-3}$. From the in situ data over a long time baseline, the background electron density on average is $\sim$ 5~cm$^{-3}$. From this, the DM excess expected due to the background SW is  $2.8\times10^{-3}$~pc~cm$^{-3}$. Hence the DM excess from the in situ data is $\sim$ $2.9\times10^{-3}$ pc/cc. This value conforms with the excess DM observed for PSR J1022+1001 on 09 August 2022 (see Section \ref{sec:cme_affected_epoch}).

\section{Analyzing a possible mode change epoch of PSR J2145$-$0750}
\label{sec:ModeChange}    
\begin{figure}[htbp]
        \centering
        \includegraphics[width=0.48\textwidth]{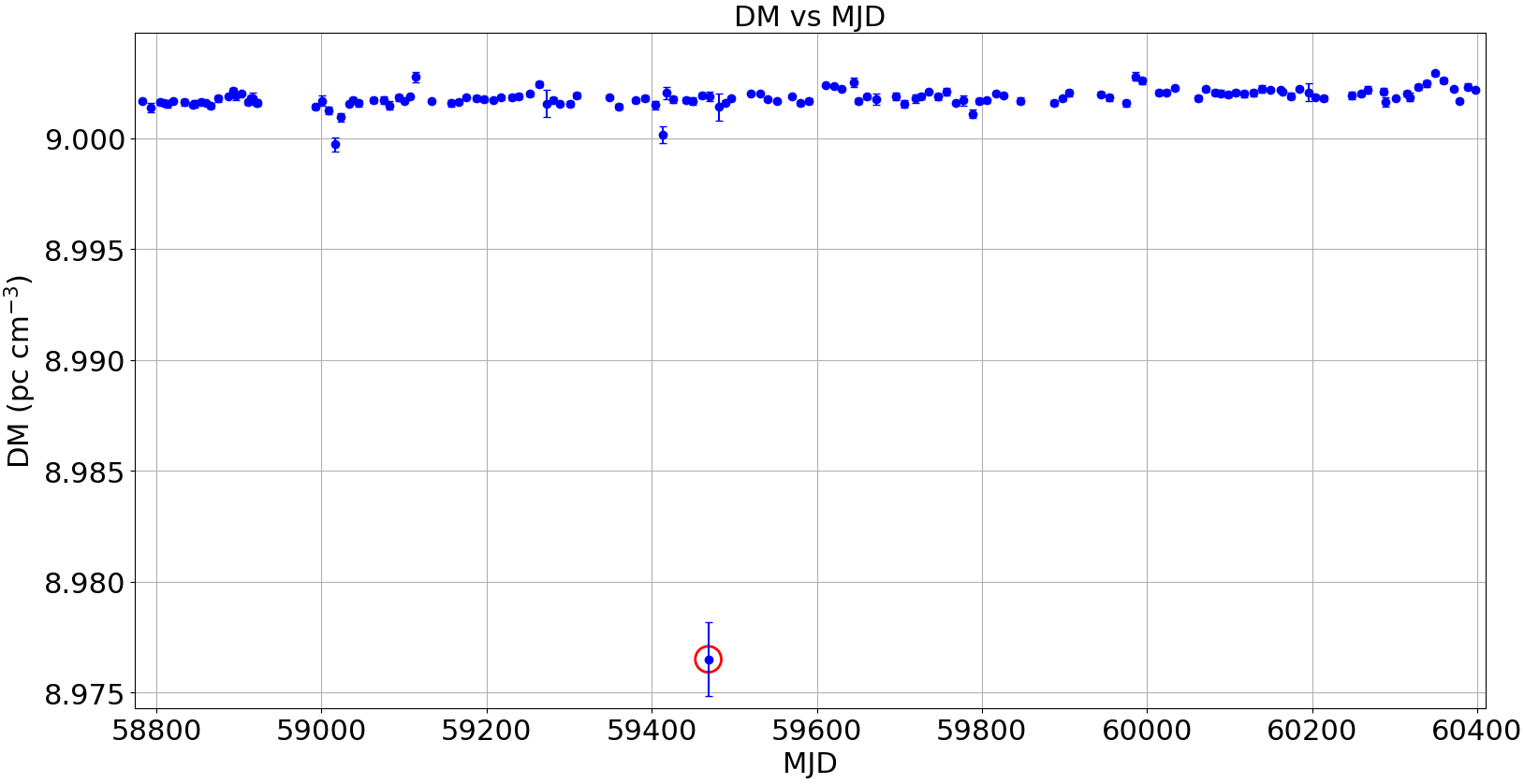}
        \caption{DM time series of PSR J2145$-$0750 along the InPTA-DR2 baseline (all prominent outliers, which are artifacts of poor data, are not included in this plot), highlighting the single outlier, corresponding to a potential mode-change, with red circle. \SC{The blue points with vertical bars represent the {\tt DMCalc}-evaluated DM with its uncertainty.}}
        \label{fig:J2145_outliers}
    \end{figure}

    \SC{Mode change in a MSP refers to the transition between two (or more) distinct pulse profile shapes, arising from a change in the configuration of the pulsar's magnetosphere~\citep{ModeChange_Magnetospheric_1, ModeChange_Magnetospheric_2}. Usually one mode (normal mode) is favoured for most of the time, and switches sporadically to alternative mode(s) (abnormal mode(s)) at other times.} \SC{For the} PSR J2145$-$0750 \SC{a normalized median absolute deviation (NMAD) outlier-search applied to its DM series identifies an outlying epoch with a very good detection and having DM deviation exceeding 3$\times$NMAD (see the red-coloured encircled point in Fig.~\ref{fig:J2145_outliers}). Besides the prominent outlier, the DM time series of PSR J2145$-$0750 shows several small time-scale deviations around its median value (see Fig.~\ref{fig:J2145_outliers}).}
    \SC{Interestingly, all of these small-time-scale deviations also correspond to} good detections with high S/N, indicating the possibility of profile showing marginal variability w.r.t. its template. To examine further\SC{, the physical significance of the prominent outlier, and the small-time-scale deviations,} we proceeded as follows.

   \subsection{Peak-ratio distribution}
   \label{sec:peakratio}
    For each observational epoch, we updated the corresponding archive/fits files using the InPTA-DR2 final timing solution. The header DM values were then corrected with epoch-wise estimates obtained from {\tt DMCalc}, and the profiles were subsequently dedispersed to remove dispersion-induced artifacts. For ready comparison and consistent analysis of all the profiles across the baseline, we positioned them at the same pulse-phase (kept the main peaks at the center). In addition, the profiles were band-equalized~\citep{Rana25} so that the power remained uniform across the frequency channels. 

\begin{figure}[htbp]
        \centering
        \includegraphics[width=0.48\textwidth]{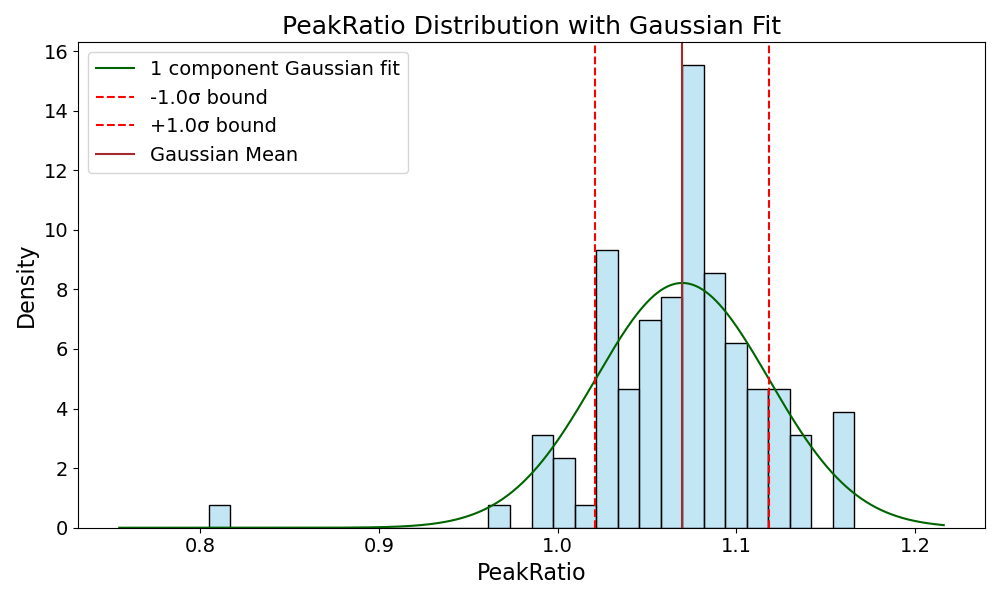}
        \caption{PSR J2145$-$0750 peak ratio distribution. The dark green curve is the Gaussian fit to the histogram of peak-ratio, with the two red vertical lines indicating $1$-$\sigma$ away from the mean peak-ratio, indicated by brown vertical line.}
        \label{fig:J2145_peakratio_distribution}
    \end{figure}

   We then quantified the mode of the pulse profile by evaluating the peak ratio, defined as the ratio of the left peak to the right peak in the two prominent components of the profile, for each epoch. The resulting distribution of the ratio of the component amplitudes is well described by a Gaussian with mean and mode value of 1.0698 and a standard deviation of 0.0486, yielding 1-$\sigma$ bounds of [1.0212, 1.1183] (see Fig.~\ref{fig:J2145_peakratio_distribution}).
   However, we identify a singular `outlier', lying outside the Gaussian envelope, with a peak ratio of $\sim$0.8, observed at MJD 59468 (11 September 2021), the same outlier seen in the DM time series plot in Fig.~\ref{fig:J2145_outliers}, indicative of a possible mode change.

\begin{figure}[htbp]
                  \centering
                  \begin{subfigure}[b]{0.36\textwidth}
                    \centering
                    \includegraphics[width=\textwidth]{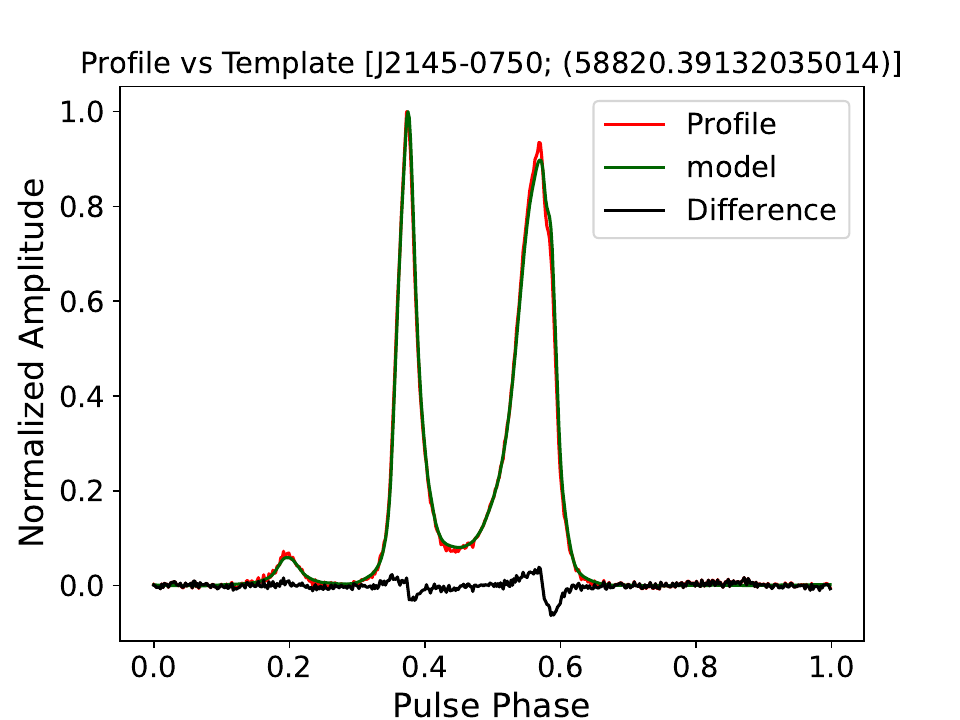}
                    \caption{58820}
                    \label{fig:J2145_58820}
                  \end{subfigure}
                  \hspace{0.02\textwidth}
                  \begin{subfigure}[b]{0.36\textwidth}
                    \centering
                    \includegraphics[width=\textwidth]{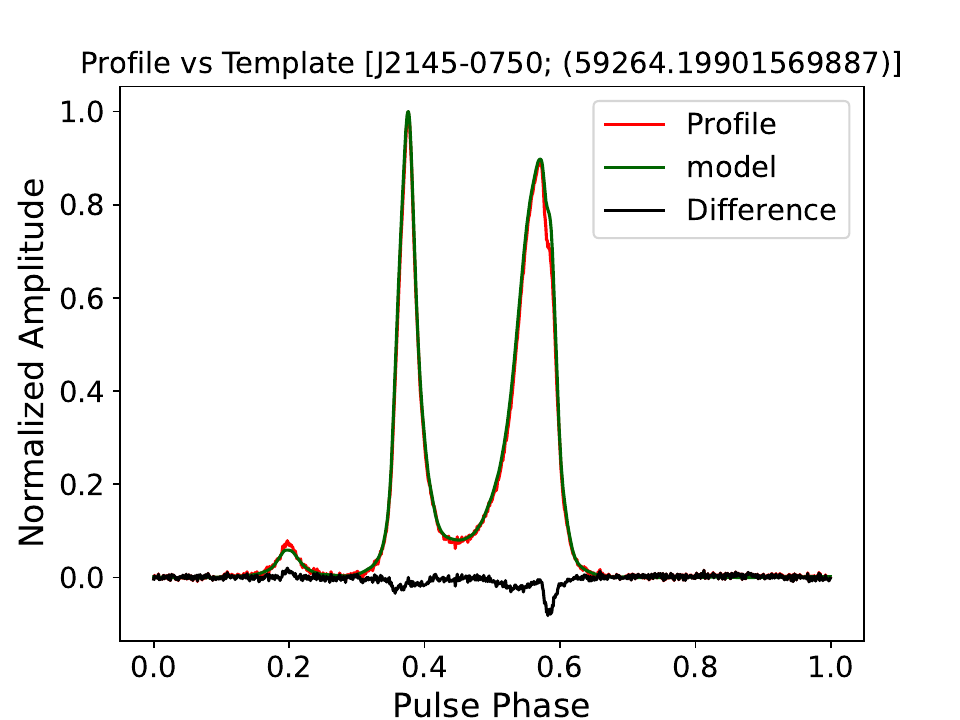}
                    \caption{59264}
                    \label{fig:J2145_59264}
                  \end{subfigure}
                  \hspace{0.02\textwidth}
                  \begin{subfigure}[b]{0.36\textwidth}
                    \centering
                    \includegraphics[width=\textwidth]{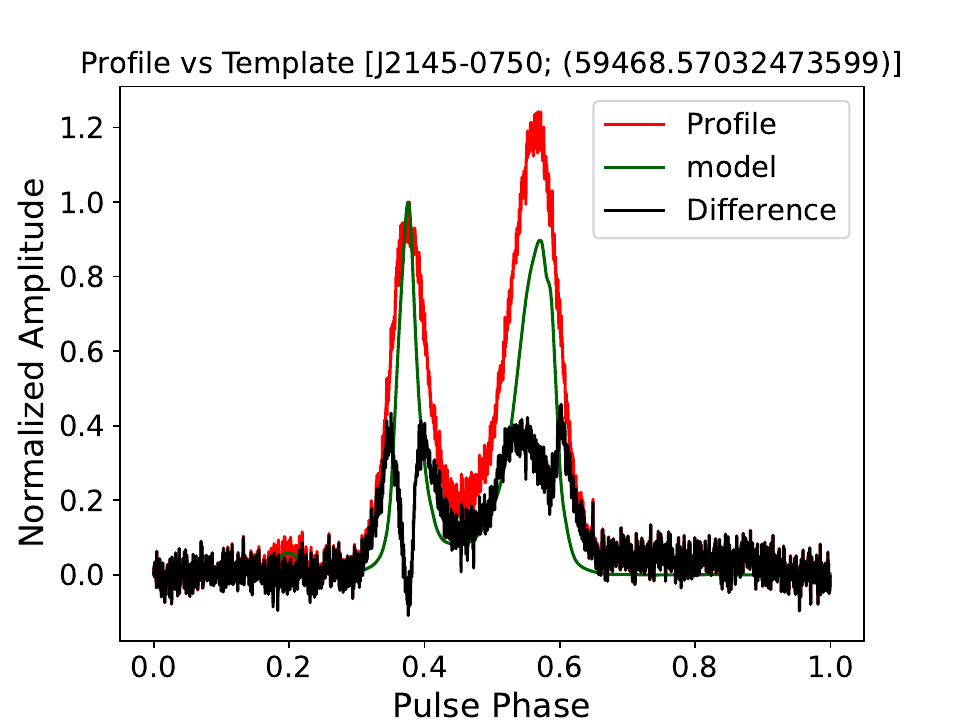}
                    \caption{59468}
                    \label{fig:J2145_59468}
                  \end{subfigure}
                  \hspace{0.02\textwidth}
                  \begin{subfigure}[b]{0.36\textwidth}
                    \centering
                    \includegraphics[width=\textwidth]{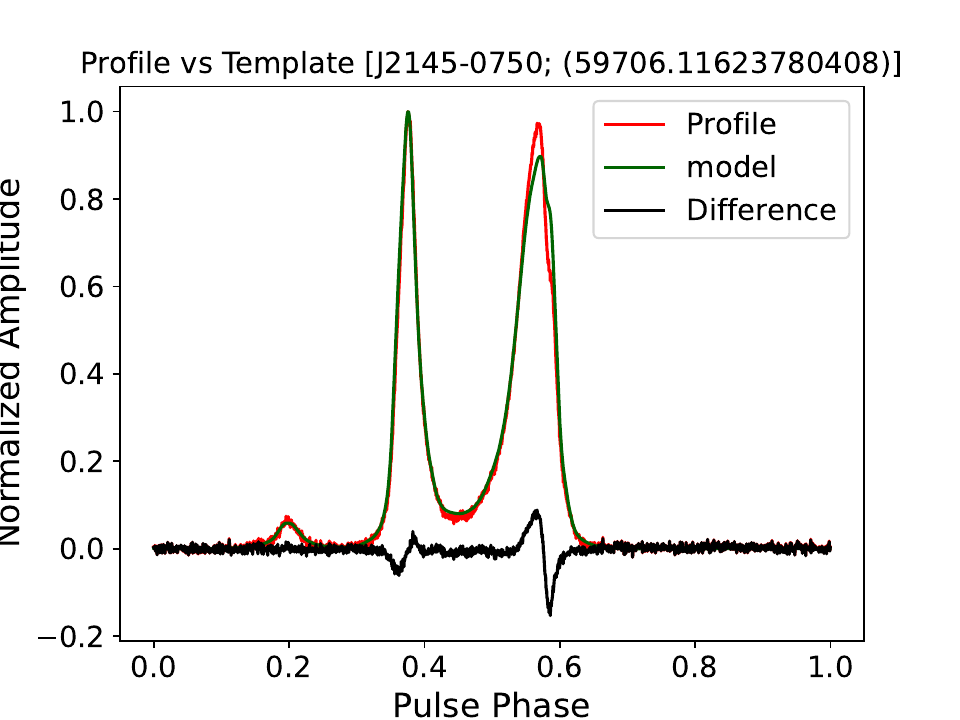}
                    \caption{59706}
                    \label{fig:J2145_59706}
                  \end{subfigure}
                  \caption{Profile comparison: The green curve in each of the above plots is the InPTA-DR2 template profile, while the red ones are the profiles corresponding to the epochs of interest, with the black curve denoting the difference between the profile and the template The left prominent peak component of each profile is normalized to unity for ready comparison.}
                  \label{fig:J2145_profilecomparisons}
        \end{figure}

    A direct comparison of the profiles across the following selected epochs confirms this interpretation (see Fig.~\ref{fig:J2145_profilecomparisons}). In particular, the profile at MJD 59468 shows distinctively different relative peak heights compared to those for MJD 58820 (close to the mode value of the peak-ratio distribution) and MJD 59706 or MJD 59264 (close to the 1-$\sigma$ bounds). These four epochs are analyzed in detail for comparative study in the subsequent discussion.

    \subsection{Profile evolution with frequency}
\begin{figure}[htbp]
        \centering
        \includegraphics[width=0.51\textwidth]{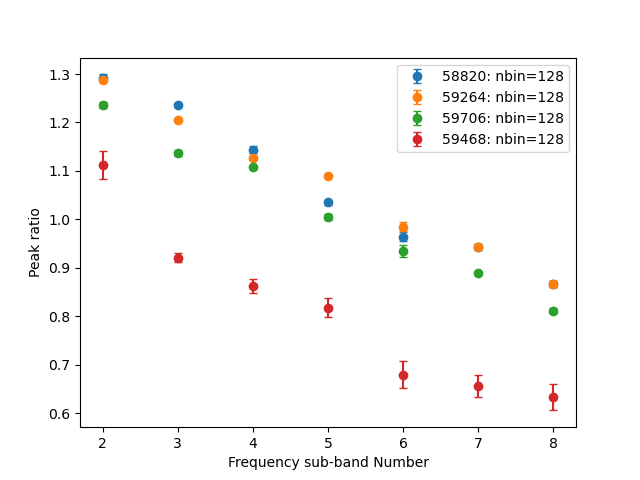}
        \caption{Profile evolution with frequency corresponding to the four epochs - MJD 58820, 59264, 59706, 59468. The X-axis corresponds to the seven out of eight subbands spanning 200 MHz band width, running from 500 MHz to 300 MHz -- subband number 1 (which is removed because of noise) corresponds to a central frequency of $\sim$486 MHz, and the subband number 8 corresponds to $\sim$311 MHz.}
        \label{fig:J2145_profevolve}
        \end{figure}

    To further test the robustness of the above result, we frequency collapsed the profiles of the four epochs into eight subbands and evaluated the peak ratio in each subband. We find that, across all epochs, the profile evolution with frequency is preserved: the left peak is smaller than the right peak at low frequencies, while the relative amplitudes gradually invert towards the higher subbands. Although the overall frequency dependence is similar, the relative peak heights at epoch MJD 59468 remain distinct from those of the other epochs, as illustrated in Fig.~\ref{fig:J2145_profevolve}. For all the four epochs in this analysis, the number of phase bins ($nbin$) have been collapsed to 128 from their initially higher number of phase bins -- this ensures reduction in the noise floor, and eliminates the possibility of such a mode change being an artifact of the radiometer noise. 

    \subsection{Eliminating systematics degenerate with mode change}
    The peak-ratio distribution reveals a single stable mode, \SC{i.e. the normal mode}, well described by a Gaussian, and a distinct outlier corresponding to the epoch MJD 59468. The latter lies outside the Gaussian envelope and therefore raises the natural question of whether \SC{it corresponds to an abnormal mode, representing} a genuine mode-change event or merely an artifact arising from \SC{propagation effects or} observational systematics. Potential sources of such artifacts include: inter-stellar dispersion, which can alter the pulse shape; inter-stellar scintillation, which modulates the pulsar signal across frequency channels and can affect the frequency-integrated profile (particularly relevant for J2145–0750, which exhibits strong frequency-dependent profile variations); polarization calibration errors or flips in polarization channels; and radiometer noise, which may distort the band shape or modify the root-mean-square (RMS) of the profile baseline.

    Each of these systematics has been carefully accounted for as follows. Dispersion effects are eliminated by using highly precise, epoch-wise DM estimates obtained with {\tt DMCalc} on dual-band observations. Scintillation is accounted for by inspecting subbanded profiles for the four epochs of interest, which averages out the variation in signal strength across frequency channels \SC{constituting each of the subbands}. Band-shape equalization and phase bin collapsing suppress the impact of radiometer noise by reducing the RMS of the noise floor. The polarization calibration related systematics would be ineffective in our dataset since they are total intensity data rather than the full four Stokes parameter recordings. It is for this reason, however, an intrinsic profile change in polarized components would remain undetected in our data. Additionally, PSR J2145$-$0750 was at a separation angle of $\sim 160 \degree$ with the Sun, at the potential mode-changing epoch MJD 59468, thereby ruling out any solar origin for this outlier. Furthermore, we checked the observatory settings and the observatory logs for that epoch MJD 59468 -- we found no mention of any observatory related glitches  or atmospheric events, that could have attributed to such a potential mode change; also the observatory settings for that day was the standard one used for InPTA observations.

    \subsection{Indication of potential intrinsic magnetospheric changes}

\begin{figure*}[htbp]
        \centering
    
        \begin{subfigure}[b]{0.49\textwidth}
            \centering
            \includegraphics[width=\textwidth]{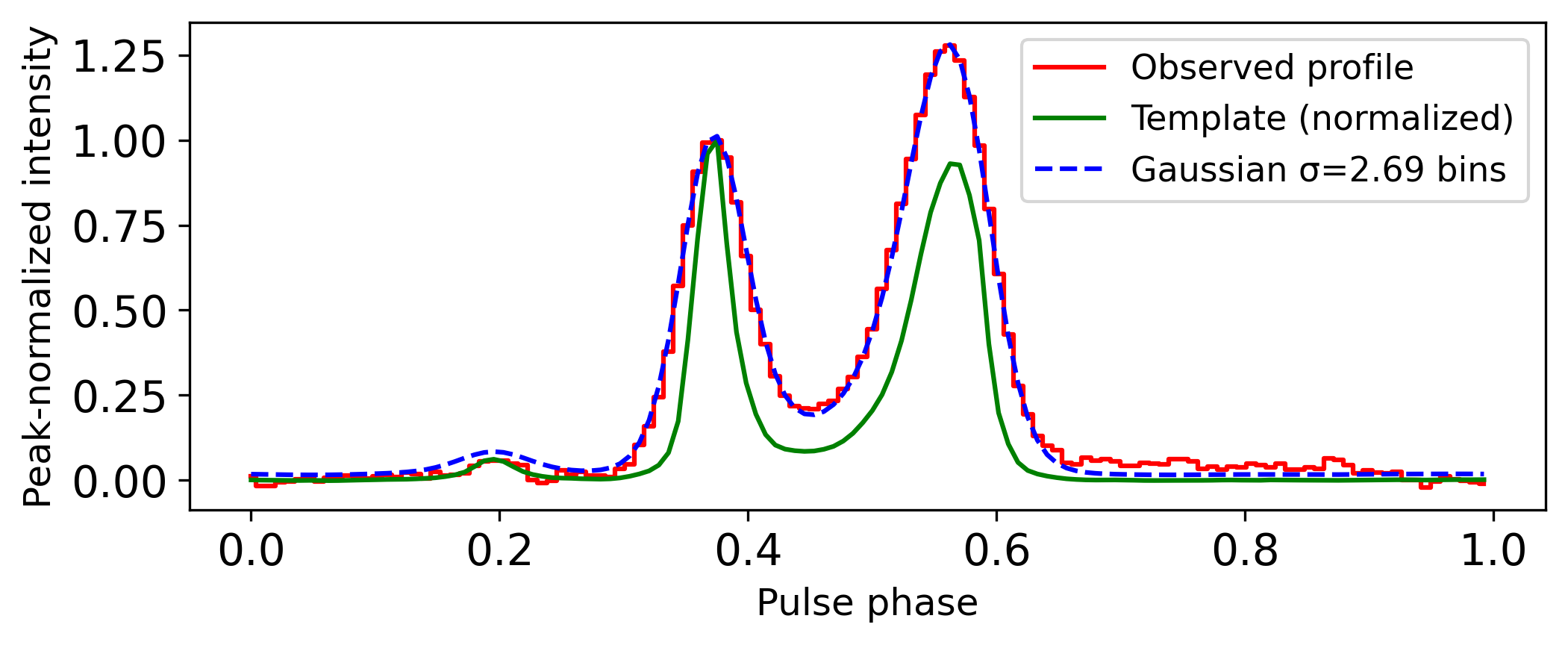}
            \caption{Frequency-integrated profile}
            \label{fig:subband1}
        \end{subfigure}
        \hfill
        \begin{subfigure}[b]{0.49\textwidth}
            \centering
            \includegraphics[width=\textwidth]{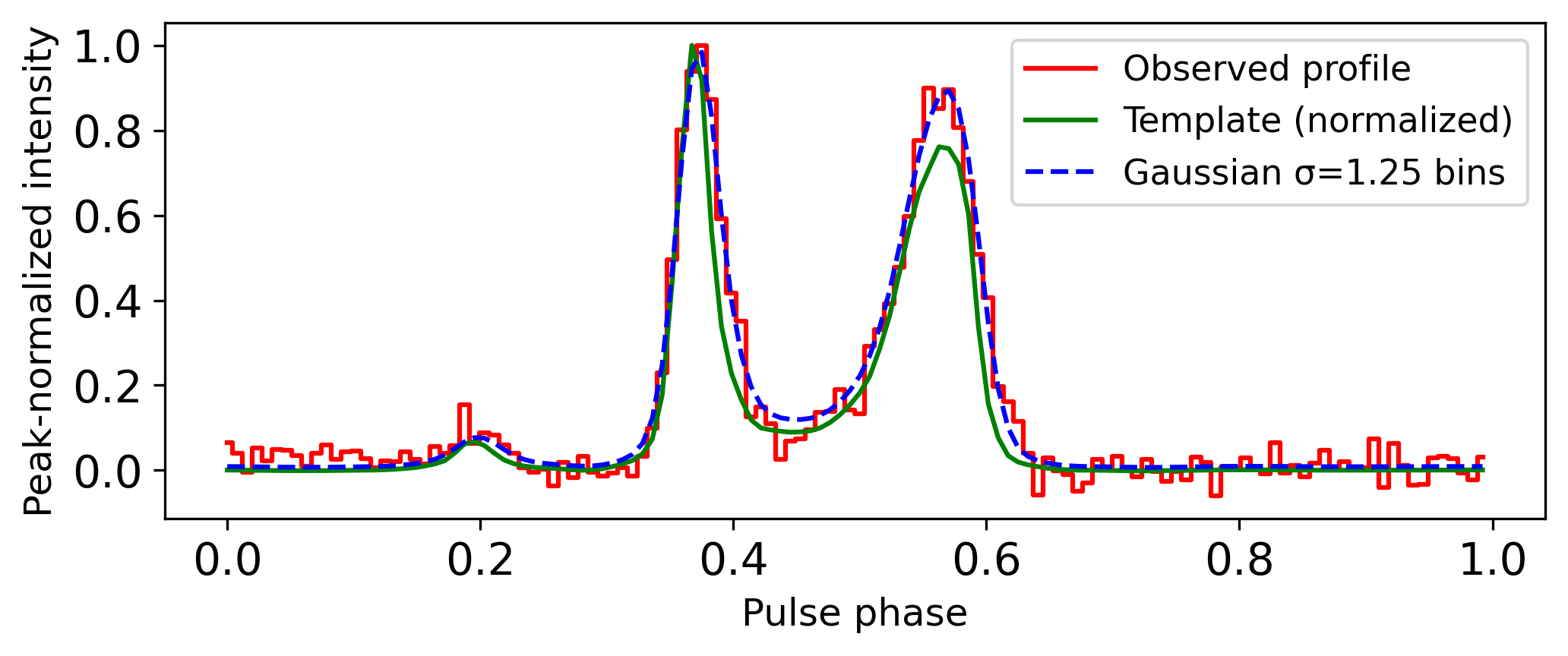}
            \caption{Subband 2 (central frequency = 461 MHz)}
            \label{fig:subband2}
        \end{subfigure}
    
        \vspace{0.01\textwidth}
    
        \begin{subfigure}[b]{0.49\textwidth}
            \centering
            \includegraphics[width=\textwidth]{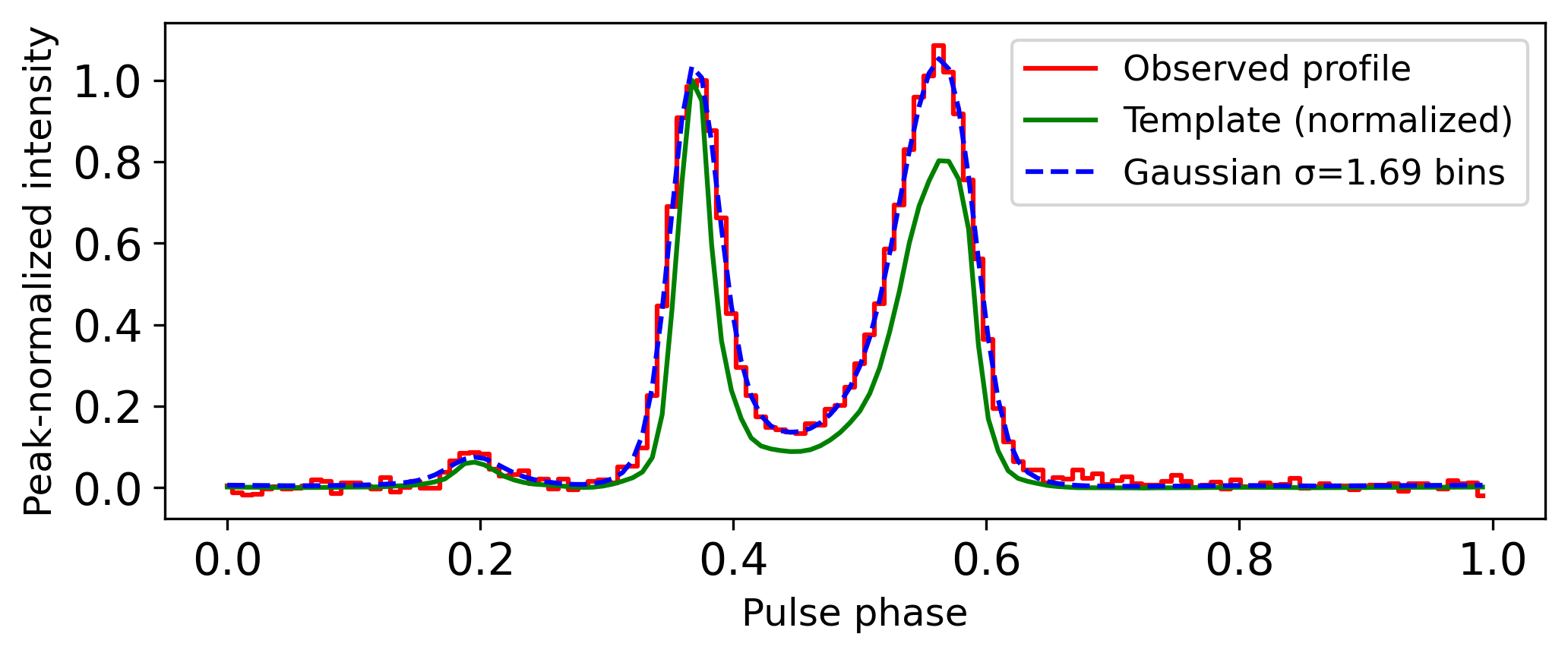}
            \caption{Subband 3 (central frequency = 436 MHz)}
            \label{fig:subband3}
        \end{subfigure}
        \hfill
        \begin{subfigure}[b]{0.49\textwidth}
            \centering
            \includegraphics[width=\textwidth]{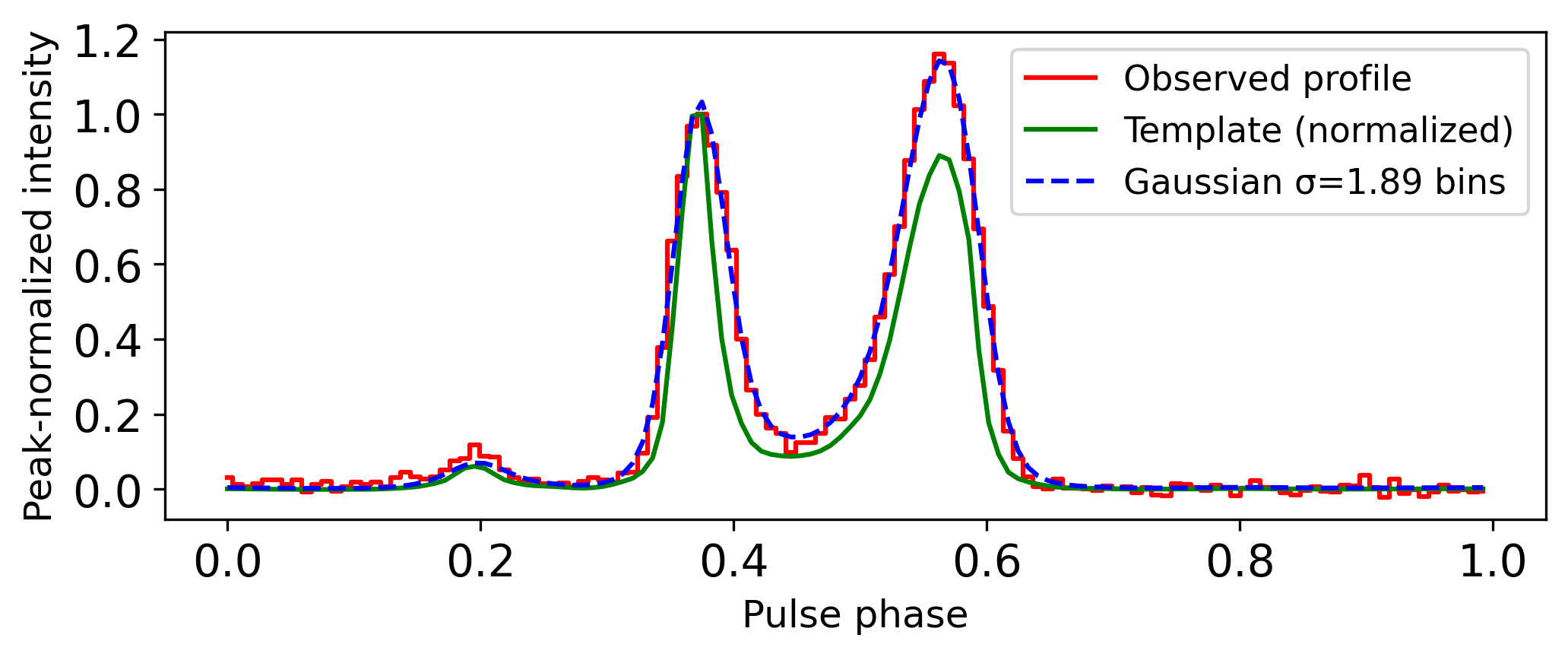}
            \caption{Subband 4 (central frequency = 411 MHz)}
            \label{fig:subband4}
        \end{subfigure}
    
        \vspace{0.01\textwidth}
    
        \begin{subfigure}[b]{0.49\textwidth}
            \centering
            \includegraphics[width=\textwidth]{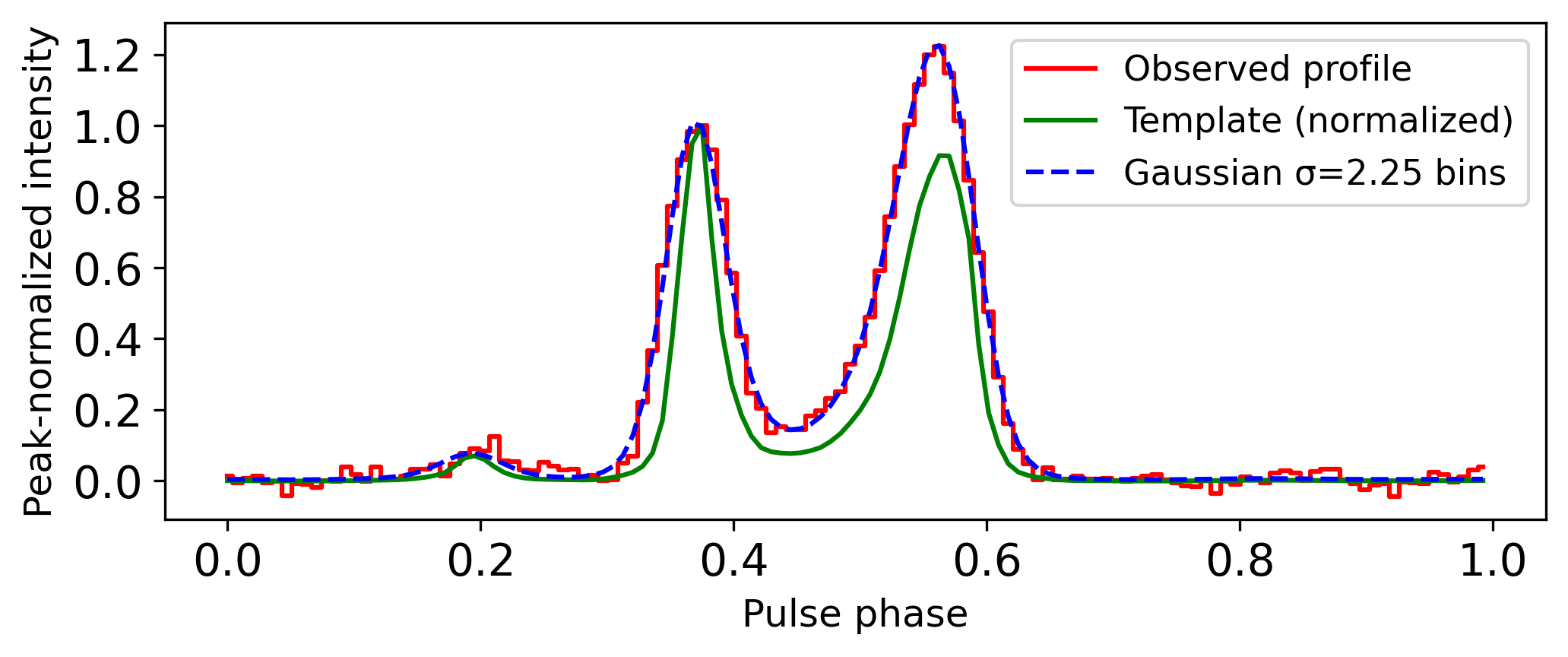}
            \caption{Subband 5 (central frequency = 386 MHz)}
            \label{fig:subband5}
        \end{subfigure}
        \hfill
        \begin{subfigure}[b]{0.49\textwidth}
            \centering
            \includegraphics[width=\textwidth]{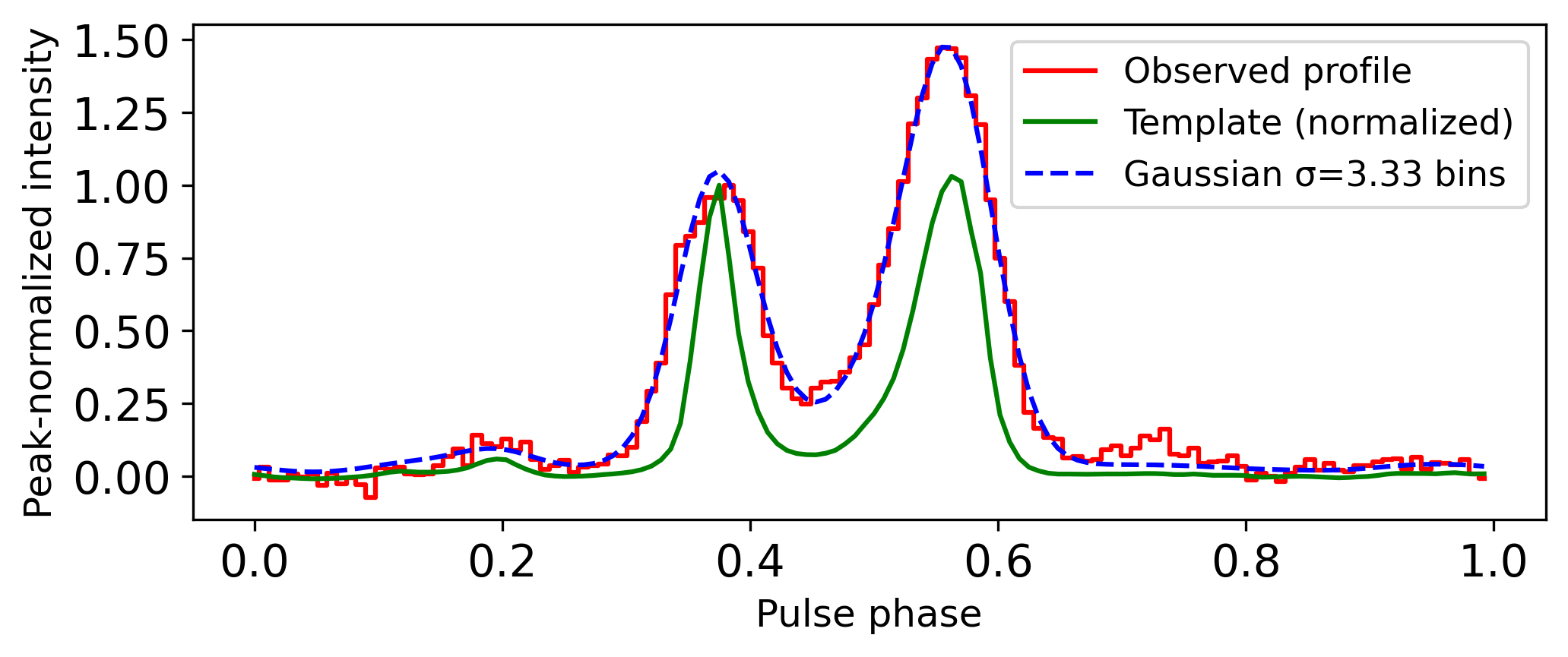}
            \caption{Subband 6 (central frequency = 361 MHz)}
            \label{fig:subband6}
        \end{subfigure}
    
        \vspace{0.01\textwidth}
    
        \begin{subfigure}[b]{0.49\textwidth}
            \centering
            \includegraphics[width=\textwidth]{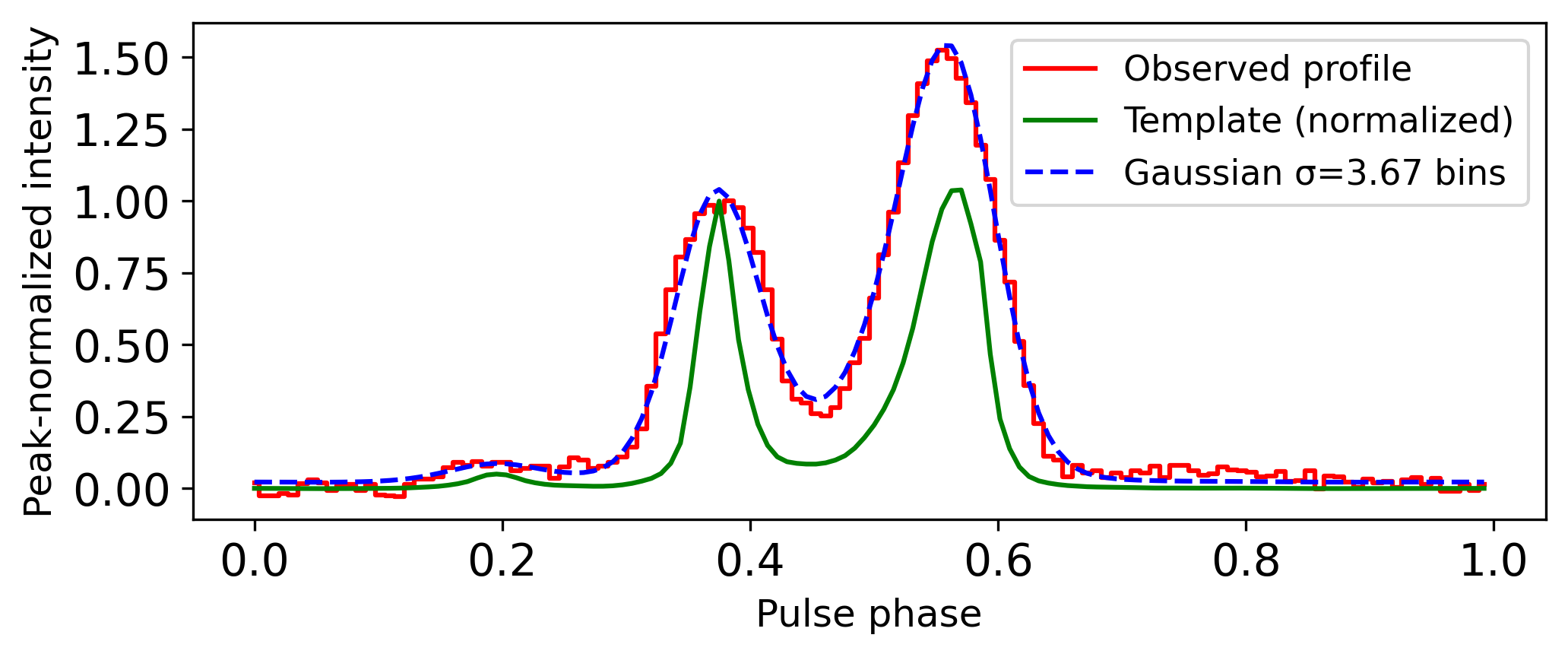}
            \caption{Subband 7 (central frequency = 336 MHz)}
            \label{fig:subband7}
        \end{subfigure}
        \hfill
        \begin{subfigure}[b]{0.49\textwidth}
            \centering
            \includegraphics[width=\textwidth]{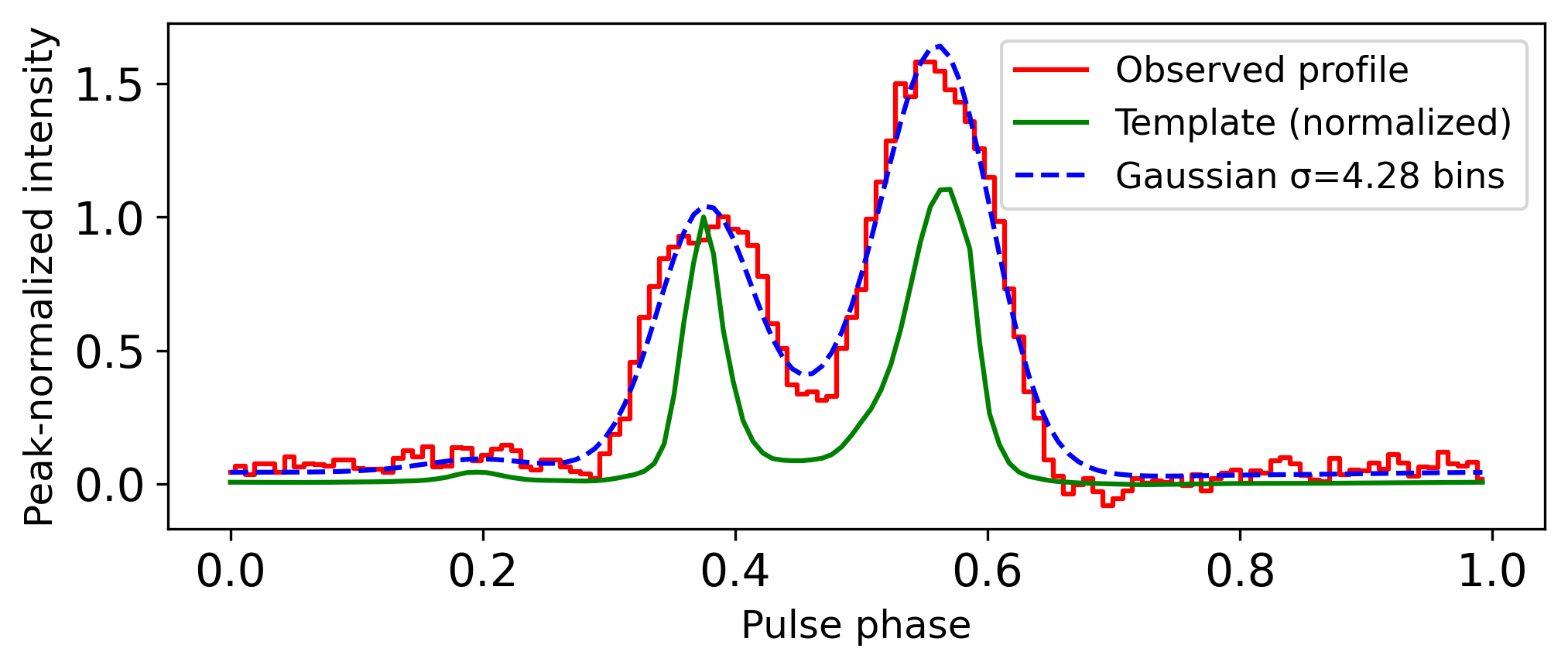}
            \caption{Subband 8 (central frequency = 311 MHz)}
            \label{fig:subband8}
        \end{subfigure}
    
        \caption{Convolution plots: The green curve in each of the above plots is the InPTA-DR2 template profile for the full band in (a) and corresponding subbands from (b)-(h), while the red ones are the profiles corresponding to the full band in (a) and subbands in (b)-(h). The blue dashed curve is the best-fit convolved template with a Gaussian kernel. Subband 1 is not considered since it has poor S/N.}
        \label{fig:convolution_59468}
    \end{figure*}

    Apart from the change in the relative heights of the peaks of the profile we also see a {\it symmetric} broadening of the pulse width. We explore this, as a final test, to establish the plausible theoretical reason behind the potential mode change. We convolve each of the subbands of the template with a symmetric Gaussian kernel and find the best fit (using $\chi^2$ test) which resembles the actual subbanded profile. \SC{The parameters used for this Gaussian convolution are $\{A,b,s,\sigma_g\}$, where $A$ is the overall amplitude scaling, $b$ is the base line offset, $s$ is the phase-shift, and $\sigma_g$ is the standard deviation.} \SC{The value of $\sigma_g$ corresponding to the best-fit Gaussian kernel, which we denote by $\sigma$\footnote{\SC{This $\sigma$ is not to be confused with the one described in Section \ref{sec:peakratio}}},} provides an estimate of the {\it observed} broadening in the corresponding subbanded profile. From the results of this analysis (see Fig.~\ref{fig:convolution_59468}), we find non-zero $\sigma$ (in bin units) for each of the subbands. 
    
    When we fit the subbanded $\sigma$s with a power law, we find its frequency dependence as $\sigma \propto \nu^{-2.91}$ \SC{($\nu$ being the central frequency of a subband)}, which is different from that due to any known propagation effect. In fact, for a non-outlying epoch, say MJD 58820 (\SC{corresponds to the} mode of the peak-ratio distribution), the absence of any such visible pulse-width broadening, comes in support for the potential mode-changing event. From the $\sigma$ value of each subband we deduce an estimate of the \SC{putative DM error\footnote{Here by DM error we refer to the difference in true DM and {\tt DMCalc}-evaluated DM.} ($\Delta DM$) that could, in principle, lead to such broadening.} \SC{To do so,} we equate the residual dispersion delay ($\Delta t_{\Delta DM}$) across a subband, \SC{arising from a putative DM error,} with the observed profile-broadening $\sigma$ for that particular subband:
    \begin{equation}
        \Delta t_{\Delta DM}~(ms)= 8.3 \times10^{6}\times \Delta DM \times \frac{\Delta \nu}{\nu^{3}}~=~\sigma~(ms) = \sigma~(bins) \times \frac{P}{nbin}~,
    \end{equation}
    where $P$ is the pulsar spin period, and $\Delta \nu$ is the width of the subband.  We observe that the estimated $\Delta DM \sim 0.08~ pc/cc$, for all the subbands, exceeds the DM uncertainty $\sim 0.001~pc/cc$, by nearly two orders of magnitude. Hence, such a huge error in DM would have been clearly visible as a parabolic drift in the frequency-phase plot of the dedispersed frequency-integrated profile, which we do not see. Therefore, the 
    source for the observed pulse-width broadening is less likely to be due to dispersion effects. Scattering effect, if any, is also possibly not the dominating contributor to such a pulse-width broadening, as validated from the absence of well-defined asymmetric exponential tails of the individual pulses in the profile and that clear scintles of 3--4~MHz wide scintillation bandwidths were seen in the observations. The symmetric broadening in the pulse-widths, therefore, in all possibilities, indicate a potential magnetospheric origin.

\section{\SC{Distinguishing CME-induced and mode-change-induced outliers}}
\label{sec:compare}

\SC{The two DM outliers investigated in this work arise from physically distinct mechanisms, and consequently required different criteria for their identification and establishing their physical origins. In this section we integrate the fundamental elements of the two case studies, explicitly comparing them for coherence.} 

\SC{In PSR J1022+1001, the identified outlier (MJD 59800) occurs at a small solar elongation and shows a clear DM excess above the modelled solar-wind contribution, with no accompanying change in the pulse shape -- the criteria of small solar elongation is essential because heliospheric plasma structures can measurably affect the dispersion measure only when the line of sight passes reasonably close to the Sun. Since CMEs influence only the integrated electron density and do not alter the intrinsic properties of the pulsar, the absence of any profile-shape deviation is an essential discriminant that rules out intrinsic variability. The DM excess is subsequently found to be consistent with a CME intersecting the line of sight of the pulsar, as supported by DONKI WEC simulations, LASCO/C3 coronagraph imaging, and STEREO-A in situ plasma data.}

\SC{For PSR J2145$-$0750, the situation is reversed. The identified outlier (MJD 59468) does not coincide with small solar elongation, but instead shows significant changes in pulse shape quantified by peak-ratio value well outside the stable distribution. Such morphology-altering behaviour at large solar elongation is indicative of a possible mode-changing event and cannot be produced by heliospheric propagation effects, as the pulsar line of sight passes far from the solar corona, rendering any CME-induced DM fluctuation negligible. This is further supported by symmetric pulse-width broadening across frequency subbands and rigorous checks against dispersion errors, scintillation, radiometer noise, and instrumental effects.}

\SC{These two case studies demonstrate that DM outliers in PTA datasets arise from distinct physical mechanisms and therefore require different diagnostic criteria. CME-induced outliers manifest as excess DM at small solar elongations without accompanying profile changes, whereas mode-change events are characterized by simultaneous DM excursions and pulse-shape anomalies.}

\section{Conclusions}
\label{sec:conclude}

\SC{In this work, we investigated astrophysically motivated outliers in the InPTA-DR2 DM time series for two representative pulsars, PSR J1022+1001 and PSR J2145$-$0750. A key aspect of our approach is that the identification of meaningful outliers in PTA datasets cannot be based solely on statistical deviation but must be guided by physical context, since different mechanisms -- such as heliospheric plasma structures or intrinsic mode change -- imprint fundamentally different observational signatures. This context-driven framework naturally leads to distinct criteria for selecting outlier epochs in the two pulsars and provides a coherent interpretation for why their observational characteristics differ.}

For PSR J1022+1001, we identified a distinct DM excursion, \SC{in comparison to the spherically symmetric solar-wind model prediction}, on MJD 59800 (09 August 2022), which is geometrically consistent with a CME event detected by LASCO/C3 coronagraph, STEREO-A satellite and the DONKI database. Quantitative comparison between the DM excess inferred from radio data and that derived from in situ solar plasma measurements revealed excellent agreement, strongly establishing one of the first direct radio signatures of a CME along this particular PTA pulsar's line of sight. This result underscores the potential for tracing heliospheric plasma dynamics and properties from the PTA datasets at astronomical distances where satellites are not present. Additionally, the number of IPTA pulsars monitored all across the sky is large enough, compared to space-based satellites, to possibly generate a three dimensional understanding of the outflow and interactions of solar wind and CMEs. The combination of several PTA datasets, across larger baselines, has the potential to contain more such solar-activity-induced outliers - this present understanding will therefore help in better modelling of  the red-noise in the PTA dataset, induced by solar activities. This will lead to more accurate timing without having to throw away a significant portion of the dataset, at low separation angles, in the absence of better solar wind modelling. 

For PSR J2145$-$0750, the analysis of high S/N profiles revealed a statistically significant outlier in the peak-ratio distribution at MJD 59468 (11 September 2021), corresponding to a clear change in the relative amplitudes and widths of the two principal pulse components. Systematic checks indicated that the broadening and altered peak ratio are possibly intrinsic to the pulsar. The observed frequency dependence of the symmetric pulse-width broadening, further supports a possible magnetospheric origin rather than propagation effects. Since our analysis is restricted to total intensity data, we cannot directly assess whether the observed mode change is associated with changes in the polarization components. 
The event likely represents a short-lived mode-change, which left its imprint on the band 3 prominently, marking the first such indication for this source within the InPTA frequency band. This study 
underscores the importance of recognizing such mode-changing behavior in, not only PSR J2145$-$0750, but also, other pulsars in the PTA datasets, and analyzing them further for deeper understanding of such transients. Studying the physical origins of such mode-changes paves the path towards better timing and thus noise modelling, without having to discard potential outliers in efforts to make the data ready for Gravitational wave background (GWB) analysis.

Together, these results highlight the dual scientific potential of PTA observations: as sensitive detectors of solar plasma phenomena and as monitors of intrinsic pulsar magnetospheric variability. The methodology developed here -- combining detection of scientific DM-outliers, analyzing solar event data, and frequency-resolved profile analysis -- can be systematically extended to the full InPTA-DR2 sample to uncover additional transient phenomena. Future work will focus on (i) expanding this framework to other InPTA-DR2 pulsars with low ecliptic latitudes to identify further CME-related DM excursions, (ii) incorporating polarization and higher time-resolution data to explore the rotation variations due to solar events as well as physical triggers of mode changes, (iii) correlating DM and profile variability, and (iv) incorporating these propagation and intrinsic effects into the timing and noise modelling for PTA science. These efforts will deepen our understanding of both the heliospheric plasma environment and the dynamic magnetospheres of millisecond pulsars, thereby improving PTA data fidelity for gravitational-wave detection.

\section*{Acknowledgements}
The authors acknowledge the use of data from the STEREO mission, obtained through the CDAWeb service of NASA’s Space Physics Data Facility (SPDF). We thank Christina Lee (UCB/SSL), Dr. Antoinette Galvin (University of New Hampshire), and the CDAWeb team for providing access to the in situ data used in this paper.
SC and MB thank the Department of Atomic Energy for its support in hosting ``Decoding Outliers and Taming Jitters: A Deep Dive into InPTA Data''  through Apex-I Project - Advance Research and Education in Mathematical Sciences at IMSc. During this workshop, a significant fraction of the work presented in this paper was carried out.
AKP is supported by CSIR fellowship Grant number 09/0079(15784)/2022-EMR-I.
AmS is supported by CSIR fellowship Grant number 09/1001(12656)/2021-EMR-I.
AS and KR are supported by UGC JRF fellowship. BCJ acknowledges the support from Raja Ramanna Chair fellowship of the Department of Atomic Energy, Government of India (RRC – Track I Grant 3/3401 Atomic Energy Research 00 004 Research and Development 27 02 31 1002//2/2023/RRC/R\&D-II/13886 and 1002/2/2023/RRC/R\&D-II/14369). BCJ also acknowledges support from the Department of Atomic Energy Government of India, under project number 12-R\&D-TFR-5.02-0700. HT is supported by DST INSPIRE Fellowship, INSPIRE code IF210656.
KV is supported  by CSIR JRF Fellowship.
KT is partially supported by JSPS KAKENHI Grant Numbers 20H00180, 21H01130, 21H04467, and 24H01813, and Bilateral Joint Research Projects of JSPS (120237710).
NDB is supported by DST-WISE fellowship (DST/WISE-PDF/PM-17/2024). SD is supported by SERB MTR/2023/000384.
ZZ is supported by the Prime Minister’s Research Fellows (PMRF) scheme, Ref. No. TF/PMRF-22-7307.

\bibliographystyle{elsarticle-harv} 
\bibliography{main2}

\end{document}